\documentclass[lettersize,journal]{IEEEtran}
\usepackage{amsmath,amsfonts}
\usepackage{array}
\usepackage[caption=false,font=normalsize,labelfont=sf,textfont=sf]{subfig}
\usepackage{textcomp}
\usepackage{stfloats}
\usepackage{url}
\usepackage{verbatim}
\usepackage{graphicx}
\usepackage{cite}
\usepackage{caption}
\usepackage{subcaption}
\usepackage{xcolor, color, soul}
\usepackage{xfrac}
\usepackage{soul} 
\usepackage[normalem]{ulem} 
\usepackage{amsthm}

\usepackage{etoolbox} 

\usepackage{subcaption}  
\usepackage{graphicx} 
\usepackage[utf8]{inputenc}
\PassOptionsToPackage{caption=false}{subfig}
\usepackage{tabularx} 
\usepackage{booktabs} 
\sethlcolor{white} 
\usepackage{amssymb}
\usepackage{algorithm}
\usepackage{algpseudocode}
\usepackage{amsmath}
\usepackage{multirow}

\newcounter{subsubsubsection}[subsubsection]
\renewcommand{\thesubsubsubsection}{\thesubsubsection.\arabic{subsubsubsection}}
\newcommand{\subsubsubsection}[1]{%
  \refstepcounter{subsubsubsection}%
  \paragraph{\thesubsubsubsection\quad #1}%
}

\usepackage{etoolbox}
\makeatletter
\pretocmd{\subsubsubsection}{\addcontentsline{toc}{paragraph}{\protect\numberline{\thesubsubsubsection}#1}}{}{}
\makeatother

 \usepackage[compact]{titlesec}         
 \titlespacing{\section}{2pt}{2pt}{2pt} 

\AtBeginDocument{%
  \setlength{\abovedisplayskip}{3pt}%
  \setlength{\belowdisplayskip}{3pt}%
}

\begin{document}

\title{ Two-Tier High Altitude Platform Stations (HAPS) for Exploring Wireless Energy Harvesting}

\author{Faicel Khennoufa, Khelil Abdellatif, Halim Yanikomeroglu,~\IEEEmembership{Fellow,~IEEE}, Safwan Alfattani,~\IEEEmembership{ Member,~IEEE}, Metin Ozturk,~\IEEEmembership{ Senior Member,~IEEE}, Ferdi Kara,~\IEEEmembership{ Senior Member,~IEEE}
\thanks{ F. Khennoufa is with Ecole Nationale Supérieure des Technologies Avancées (ENSTA), Department of Basic Training in Computer Science, Algiers, Algeria, email: faicel.khennoufa@ensta.edu.dz. A. Khelil is with LGEERE Laboratory, Department of Electrical Engineering, Echahid Hamma Lakhdar University, El-Oued, Algeria, email: abdellatif-khelil@univ-eloued.dz. H. Yanikomeroglu is with Non-Terrestrial Network (NTN) Laboratory, Department of Systems and Computer Engineering, Carleton University, Ottawa, K1S 5B6, ON, Canada, email: halim@sce.carleton.ca. S. Alfattani is with King AbdulAziz University, Saudi Arabia, email: smalfattani@kau.edu.sa.
M. Ozturk is with Electrical and Electronics Engineering, Ankara Yıldırım Beyazıt University, Ankara, Turkiye, email: metin.ozturk@aybu.edu.tr. F. Kara is with Ericsson AB, Sweden (e-mail: ferdi.kara@ericsson.com)

}
}


\maketitle
\begin{abstract}
In sixth-generation (6G) cellular networks and beyond, aerial platforms, such as uncrewed aerial vehicles (UAVs) and high-altitude platform stations (HAPS), are anticipated to play a crucial role in enhancing connectivity, expanding network coverage, and supporting advanced communication services. 
However, the deployment of energy-efficient onboard communication systems is essential for their widespread adoption and effectiveness. 
The integration of energy harvesting (EH) into aerial platforms is envisioned to be pivotal in promoting both energy and cost efficiency. 
In this paper, we propose a new paradigm for aerial platforms in which they can collect energy from the transmitted signals of nearby aerial platforms. 
The paper employs a two-tier architecture with HAPS super-macro base stations (HAPS-SMBS) system: regular HAPS-SMBS nodes serve as base stations, while a "mother" HAPS-SMBS node acts as a manager to coordinate communications between regular HAPS-SMBS and the ground station, thus enabling wireless energy transfer.
Specifically, we analyze the characteristics of EH-enabled HAPS-SMBS and compare their performance with those without EH. Additionally, we derive the optimal regular HAPS-SMBS positioning to mitigate signal attenuation and power loss. 
Subsequently, we formulate a joint optimization problem for regular HAPS-SMBS positioning and the EH factor. We solve the problem using the iterative distance and EH factor algorithm (IDFA); however, we employ $Q$-learning to verify its effectiveness.
Our findings indicate that, compared to conventional EH systems, IDFA and $Q$-learning exhibit higher data rate performance. In contrast, $Q$-learning outperforms IDFA systems in linear models with intensive training in approximating optimal values.
Furthermore, maximizing transmit power achieves higher gains than systems without EH.
\end{abstract}


\section{Introduction}
Amidst the ongoing deployment of commercial fifth-generation (5G) cellular networks by network operators, scientists are actively discussing the emerging use cases, the prerequisites, and the latest technology that will enable the sixth-generation (6G) networks \cite{11168452}. The Third Generation Partnership Project (3GPP) has started the standardization process for 5G-Advanced, to deliver higher data rates, lower latency, increased capacity, better spectrum utilization than its predecessors, sustainability, high-precision positioning, improved mobility and time-critical support, alongside leveraging artificial intelligence and machine learning to boost network performance \cite{iqbal2023empowering,lin2025bridge,11168452,3gpp2019solutions,10981514}. Despite the advancements promised by 5G networks, there are still areas where terrestrial networks struggle to provide connectivity due to various natural conditions and geographical limitations. To tackle these challenges, research on 6G communication networks is shifting its focus to non-terrestrial networks (NTN), encompassing uncrewed aerial vehicles (UAVs), high-altitude platform stations (HAPS), and satellites. 6G envisages a monolithic architecture in which non-terrestrial stations complement terrestrial infrastructures, in contrast to previous generations of wireless networks that were typically restricted to terrestrial networks \cite{giordani2020non,kement2023,abbasi2024haps}.

Despite this burgeoning enthusiasm and heightened interest in aerial platforms within the realm of upcoming generations of wireless communications, several challenges persist. The most significant of these is the problem of energy scarcity, which severely limits the mission's duration. Numerous solutions have been proffered to address this limitation. One notable example is the integration of reconfigurable intelligent surfaces (RIS) into aerial platforms \cite{alfattani2021link,thapliyal2024energy,jiang2024rf,khennoufa2024error,alfattani2023resource}. RIS is composed of a thin and flexible metasurface material containing passive circuitry that controls the propagation of electromagnetic waves, thereby turning the wireless channel into a controllable environment and improving overall system performance \cite{11029408,10978385}.
Although this integration has yielded good outcomes, it is imperative to explore additional solutions as well.
On a parallel trajectory, the concept of energy harvesting (EH) has garnered substantial attention as an eco-friendly and sustainable avenue to prolong the operational lifespan of energy-constrained wireless networks, as highlighted in previous studies \cite{jiang2024rf,na2021uav,khan2024rf,monagle2025energy}. \hl{EH refers to capturing energy from external sources, such as solar radiation or ambient radio-frequency (RF) signals, and converting it into usable electrical power, providing a sustainable method for powering electronic devices. Although solar EH offers high energy potential, its effectiveness is strongly affected by adverse weather and day–night cycles.}
Hence, our attention shifts towards exploring wireless EH as an alternative. In this paper, we focus on wireless EH to extend battery life, power sensors, and support communication systems.


\subsection{Related Work}
Much research has studied the topics of RIS and EH on a larger scale, and assessed their efficacy across various applications such as multiple-input multiple-output (MIMO) and massive MIMO through evaluating key performance indicators. In this context, our focus will be on EH aspects within NTNs. 
In~\cite{8888216}, the EH and coverage probability of a two-hop aerial base station when it harvests energy from a terrestrial base station are investigated. In~\cite{9860882}, the deployment of aerial base stations is optimized to collect electromagnetic energy from ground base stations and information transmission. The authors of~\cite{9805770} demonstrate how energy efficiency and overall system performance can be enhanced by optimizing the UAV trajectory and antenna configuration to serve ground devices via the NOMA system. The UAV-RIS was used to collect energy and reflect the incident signals by adjusting its phase shifts and reflection coefficients (called EH-assisted UAV-RIS scheme) to reduce the maximum amount of energy consumed \cite{zargari2022user}. Moreover, the authors of \cite{ouamri2023joint} considered wireless EH for charging UAVs by applying multi-beam energy. To improve energy efficiency, a novel scheme was proposed in \cite{peng2023energy}, where the UAV-RIS is integrated with EH. The study in \cite{dang2022secure} examined the secrecy outage probability of UAV-assisted amplify-and-forward (AF) with EH. The airborne MIMO system was proposed in \cite{ma2022energy}, where an EH algorithm is used for ground sensors, to establish a consistent connectivity. In these studies, the results demonstrate that EH-assisted wireless systems outperform their non-EH counterparts. 


It is worth noting that the current research in the field of EH in wireless communication networks has primarily focused on using UAVs to gather power from ground base stations or ground nodes \cite{ouamri2023joint,peng2023energy,dang2022secure}. Also, through a more comprehensive vision about using EH from ground stations, going back to the past years, specifically to the 1980s, the energy beams from the ground were used to supply energy to HAPS systems using microwave beams \cite{CRC,brown1986microwaver,schlesak1988microwave}. In \cite{MyArrow}, in collaboration with researchers from the University of Tokyo, the project was carried out at the Georgia Institute of Technology to collect energy at 6.5 km from a television broadcast station in Tokyo, Japan. Moreover, National Aeronautics and Space Administration (NASA)-sponsored experiments yielded 34 kW of S-band microwave energy, with an average aggregate conversion efficiency of 82.5 $\%$ \cite{brown15866design,brown1983design}. In \cite{dickinson1976performance}, demonstration research was carried out to show that EH technology can supply electrical power levels that are sufficient for operating large-payload stratospheric vehicles.
\hl{The latest developments in RIS have enormous potential to improve energy transfer and communication in HAPS-supported networks. A multi-layer transmitting/reflecting RIS can concentrate electromagnetic waves on user equipment, thus improving EH efficiency} \cite{10947296}. \hl{An absorptive RIS can simultaneously control the phase and amplitude of each element, making it possible to selectively absorb interference and amplify the desired signal}~\cite{11305158}. \hl{The absorbed signal can be redirected, detected, or dissipated, thus providing better interference mitigation and signal quality than traditional RIS. Furthermore, recent studies have demonstrated the benefits of RIS-assisted NTNs, for example,}
Hoa T. Le \textit{et al}. proposed an NTN that integrates HAPS with RIS and UAV for data transmission and EH simultaneously from around 18-25 km. It demonstrates that the collection of EH depends on system parameters like RIS size, transmitted power, power split ratio, and elevation angle~\cite{10683508}. The authors of~\cite{10520169,10437169} proved that the simultaneous transmission of information and power from the ground with the help of RIS overcomes the fading effects of long-range HAPS links and fully exploits the degrees of freedom of RIS for EH design.


On the other hand, by leveraging artificial intelligence, researchers have focused their attention on reinforcement learning, a method that seeks to maximize predicted long-term benefits. A widely employed reinforcement learning algorithm is $Q$-learning, which utilizes value iterations. $Q$-learning can identify the optimal strategy if all available actions are consistently sampled across all states of the decision process~\cite{zhai2021q}. Therefore, artificial intelligence is anticipated to enhance the overall system performance and boost the amount of harvested energy by intelligently analyzing channel conditions, and mitigating adverse effects. There has been some work on the UAV-assisted EH with artificial intelligence, which is discussed as follows. By leveraging artificial intelligence, the authors in \cite{ouamri2023nonlinear} used multi-agent deep reinforcement learning as an approach to maximize throughput and energy efficiency. In order to maximize the system energy efficiency and ensure the quality of service requirements against jamming attacks, the authors in \cite{yang2024energy} examined a UAV-RIS system powered by EH technology in a maritime communication network under the control of a malicious jammer. In \cite{liu2023joint}, a UAV utilizes wireless power transfer to enhance mobile edge learning and overcome the limited transmit power of the Internet of Things devices. The authors in \cite{jabbari2024energy} used multi-agent deep reinforcement learning to maximize energy efficiency and minimize consumption of collected energy using solar and wireless EH techniques to power the Internet of Things and UAV stations.




Based on the previous works discussed above, there has also been interest in exploiting RF EH from ground stations \cite{ouamri2023joint,peng2023energy,dang2022secure,CRC,brown1986microwaver,schlesak1988microwave,MyArrow,brown15866design,brown1983design,dickinson1976performance}. This may be one of the solutions to supply aerial platforms with energy to extend flight time and service duration. However, focusing on collecting power from ground stations only, especially in the event of natural or other disasters, may cause communication interruptions. This requires proposing other solutions. One promising method is to utilize the electromagnetic waves that are present in the atmosphere during HAPS transmission operations. 
In the midst of HAPS transmission operations, electromagnetic waves are ubiquitous in the air, offering a plentiful energy source that is barely exploited, which provides a tremendous opportunity to collect this energy and use it effectively. Even if a low amount of RF energy is collected over very long distances, it can still be utilized to operate some low-power circuits on aerial platforms.


\subsection{Contributions}

The above studies show that EH in HAPS and UAV systems is not only possible, but can be greatly improved through optimal design~\cite{ouamri2023joint,peng2023energy,dang2022secure,CRC,brown1986microwaver,schlesak1988microwave,MyArrow,brown15866design,brown1983design,dickinson1976performance,10683508,10520169}. For example,~\cite{8888216} shows that determining the optimal distances between NTN stations (e.g., HAPS and UAVs) is crucial to maximizing coverage and EH efficiency. Long distances may still be a major problem for adequate EH; however, several solutions have been proposed to mitigate this problem, including the use of massive MIMO or advanced massive MIMO and RIS, which enhance energy efficiency~\cite{10683508,10520169,zargari2022user,peng2023energy,dang2022secure,ma2022energy,dickinson1976performance}. The use of free-space optics (FSO) also provides the ability to cope with path length and weather conditions and harvest more energy~\cite{10683508}. The studies show that these designs can significantly enhance EH efficiency even in complex environments. 
Moreover, through another perspective to collect energy amidst the proliferation of aerial platforms at different altitudes, certain aerial platforms can gather energy from other platforms when they are located within the transmission range.
To the best of our knowledge, the possibility of utilizing the RF EH from the air has not been explored yet in the context of aerial platforms. In this paper, we propose a novel paradigm to expand the scope of EH for aerial platforms beyond the ground stations, when the HAPS acts as a base station, referred to as the HAPS super-macro base station\footnote{HAPS has been envisioned as a traditional International Mobile Telecommunications (IMT) base station, as an SMBS to serve different types of users in mega-cells alongside legacy macro-cells and small-cells. Referred to as a  ``HAPS IMT base station (HIBS)" by the International Telecommunication Union (ITU), it is positioned at an altitude of around 20 km in the lower stratosphere layer \cite{ITU20233,weissberger2021itu}.} (HAPS-SMBS) \cite{do2021user}. This approach aims to harness HAPS-SMBS for the free electromagnetic energy available in the air, providing an additional energy supply chain. In particular, when HAPS-SMBS are unable to harvest energy from ground energy sources due to factors, such as natural disasters, wars, lack of line of sight (LoS), etc., the necessity arises to recharge the HAPS-SMBS each time they run out of energy. Consequently, this leads to increased operational downtime. 
Offering additional energy sources such as solar energy, wireless EH, etc., can greatly increase the capabilities of HAPS-SMBS. Wireless EH offers an alternative source of power for HAPS-SMBS that are unable to collect power from sunlight.





\hl{In the context of the practical implementation of the HAPS system, the major challenges include energy constraints, long-distance propagation, payload, and environmental factors, which affect the channel. In addition, the solar energy availability may be variable, depending upon the cloud cover and day/night cycles, which further motivates the adoption of complementary wireless EH techniques} \cite{alfattani2023multimode,alam2021high,alzenad2018fso,kurt2021vision}. \hl{By incorporating the EH system in the HAPS-SMBS, the propagation of the signal is quite different compared to the terrestrial system, as the HAPS system is specifically designed for the stratospheric channel. Therefore, it is necessary to analyze the feasibility and benefits of the EH-based HAPS-SMBS system through the modeling of large-scale attenuation and radio-frequency to direct-current (RF-to-DC) conversion. The major factors affecting the EH for the HAPS-SMBS system include the distance between the two nodes, the EH factor, non-linear EH, antenna gain, and the operating~frequency.} 

In this paper, we aim to provide a propagation model analysis for EH-enabled HAPS-SMBS communication systems. In practice, numerous experiments for real-world EH circuits have demonstrated the non-linearity of their input-output characteristics. Therefore, the non-linear EH characteristic is taken into account in this work. 
We aim to design a HAPS-SMBS-assisted EH system, and obtain the optimal HAPS-SMBS positioning, and maximum collected energy to improve the system's performance and minimize signal loss. Hence, we define a joint non-convex optimization problem for HAPS-SMBS positioning and the EH factor. It is worth noting that it is difficult to solve this problem analytically because of its non-convex nature. We propose an iterative distance and EH factor algorithm (IDFA) by splitting the problem into two subproblems. However, without dividing the problem, $Q$-learning is used to verify the effectiveness of the IDFA approach.
Finally, numerical results are provided to support the proposed propagation model. The contributions of the paper are highlighted as follows:

\begin{itemize}
\item We propose a new paradigm for aerial platforms by employing wireless EH in aerial platforms. 
We use a two-tier HAPS-SMBS system containing regular HAPS-SMBS nodes serving as base stations, alongside a mother HAPS-SMBS node acting as a manager to coordinate and manage communications between regular HAPS-SMBS and the ground station, thus enabling wireless energy transfer.
This novel approach aims to extend the duration of flight and improve the energy efficiency of HAPS-SMBS.


\item We delve into EH within practical scenarios (i.e., linear and non-linear EH). Our investigation scrutinizes the potential of using EH with HAPS-SMBS, offering insights into its applicability and efficacy in real-world environments. As a benchmark, we compare our system with those without EH schemes.

\item We investigate the propagation model analysis for EH-enabled HAPS-SMBS networks. In such dynamic environments, minimizing signal attenuation and power loss are of utmost significance, and thus we derive the optimal positioning for regular HAPS-SMBS empowered by EH.



\item  We propose a joint optimization for the regular HAPS-SMBS positioning and EH factor, aiming to maximize the data rate of the system. This problem is non-convex, making it more difficult to find the best solutions. We propose an IDFA by splitting the problem into two subproblems. However, although this method provides a suboptimal solution to the problem, we want to verify the results through joint optimization without resorting to dividing the problem into two subproblems as in~\cite{zhong2022deep}. Therefore, we propose $Q$-learning to explore a sub-optimal solution and verify the effectiveness of the IDFA approach. Thus, we develop a $Q$-learning algorithm for joint optimization of the positioning of the regular HAPS-SMBS and EH factor to maximize the data rate of the system. 



\item We employ two transmission cases in this paper. The first case occurs when the collected energy is sufficient to transmit signals, while the second situation occurs when regular HAPS-SMBS cannot collect adequate energy for signal transmission. Therefore, we suggest using a portion of the regular HAPS-SMBS stored energy to meet the required transmit power. With this approach, our objective is to minimize the amount of energy consumed from the regular HAPS-SMBS inventory. To achieve this, we propose a convex optimization problem that aims to minimize the energy consumed from the regular HAPS-SMBS inventory based on the harvested energy levels. 



\end{itemize}

\subsection{Organization of the Paper}

The rest of the paper is organized as follows. The proposed propagation system architecture and modeling for the EH-enabled HAPS-SMBS are introduced in Section II. We derive the optimal positioning for regular HAPS-SMBS empowered by EH in Section III. 
\hl{The joint optimization of regular HAPS-SMBS positioning and EH factor using IDFA and $Q$-learning framework} is given in Section IV, while Section V provides the transmit power maximization for the regular HAPS-SMBS and the quantity of EH throughout the flight mission, followed by the presentation of the simulation results in Section VI. Finally, Section VII concludes the paper and presents the challenges of future works.


\section{System Architecture and Modeling}
\hl{In this section, we introduce the architecture and system model for the proposed EH-assisted HAPS-SMBS. Specifically, we describe the proposed system and its architecture. Then, we present the EH and propagation models used in this analysis.}

\subsection{Generic System Architecture}
This subsection introduces the generic system architecture of wireless EH for HAPS-SMBS. 
As shown in Fig. 1, the system contains one HAPS-SMBS referred to as the mother HAPS-SMBS and $K$ regular HAPS-SMBS. We consider that the mother HAPS-SMBS is located on top of the regular HAPS-SMBS nodes, and operates as a central point responsible for the backhaul\footnote{In a backhaul network, the mother HAPS-SMBS acts as a manager, coordinating and managing communication between the regular HAPS-SMBS and the ground station. Signals exchanged between HAPS-SMBS nodes are primarily for network management and coordination within the backhaul system, and should not be forwarded to ground users~\cite{alam2021high,alzenad2018fso}.} links of all regular HAPS-SMBS and the ground station as referred to in \cite{alfattani2023multimode,alam2021high,alzenad2018fso,kurt2021vision}. 
The main role of the mother HAPS-SMBS is to manage information transfer between all regular HAPS-SMBS nodes, enabling the wireless energy transfer to all regular HAPS-SMBS nodes. Unlike regular HAPS-SMBS, mother HAPS-SMBS does not perform radio access network (RAN) services for terrestrial users.
\hl{It is assumed that the mother HAPS-SMBS has a larger size than regular HAPS-SMBS, allowing it to carry approximately 2–3 times more solar panel area and a proportionally larger onboard energy storage capacity} \cite{alfattani2023multimode,alam2021high,alzenad2018fso,kurt2021vision}. In contrast to mother HAPS-SMBS, the main responsibility of regular HAPS-SMBS is to perform RAN services, which involve providing communications services to ground users in different areas.
\hl{In this paper, we assume that the direct link between the mother HAPS-SMBS and ground users is neglected due to the much longer distance and higher path loss. 
To the best of the authors’ knowledge, the two-layer HAPS-SMBS architecture with wireless EH from aerial platforms has not been investigated in the literature. This framework represents the first contribution of the present work.}



\begin{figure*}[]
    \centering
    \includegraphics[width=1\columnwidth]{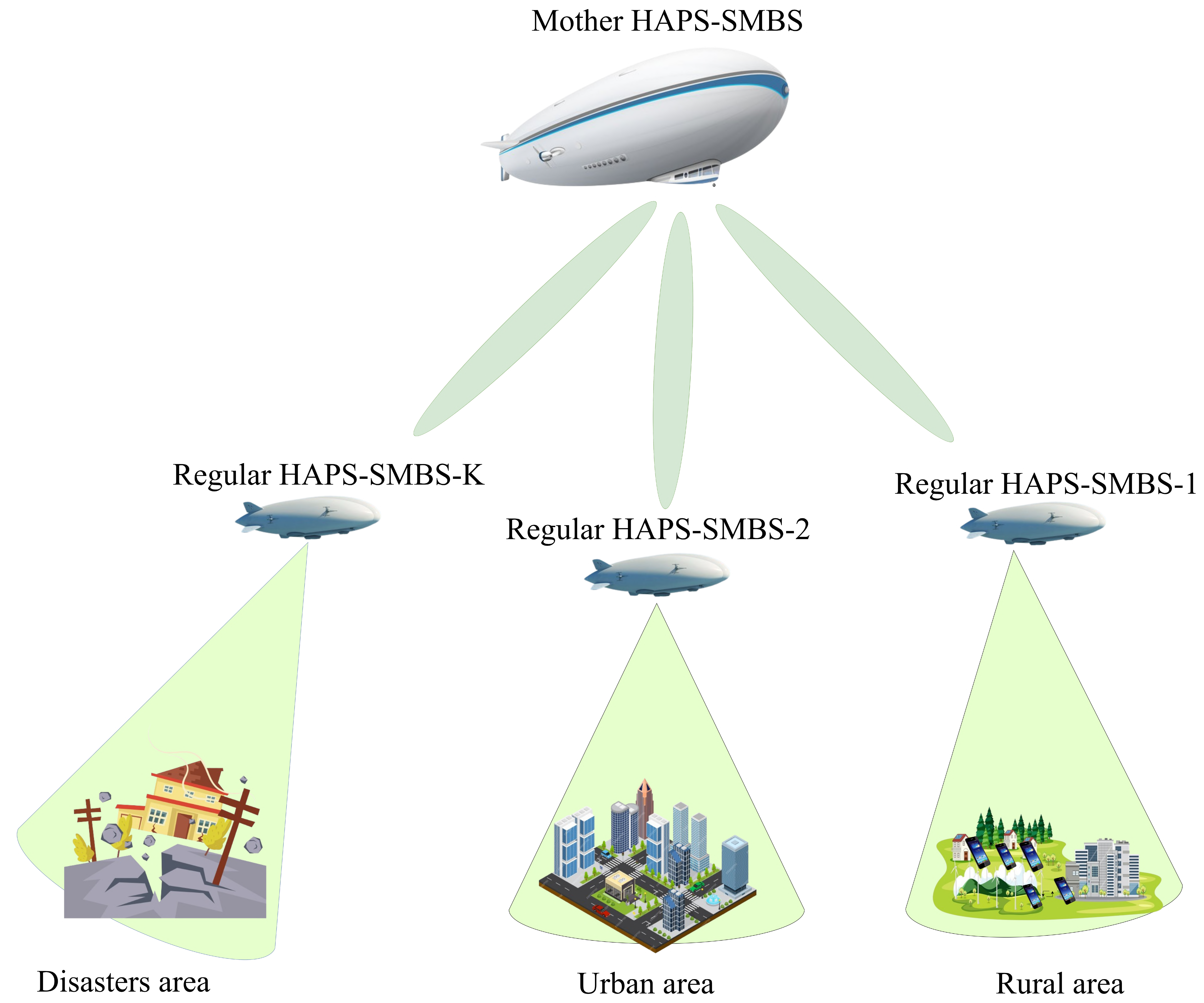}
    \caption{Generic architecture of HAPS-SMBS supported by the EH scheme.}
    \label{constellations}
\end{figure*}

\subsection{System Model}
In this paper, we evaluate a particular area, i.e., the coverage area of one regular HAPS-SMBS node as shown in Fig. 2.
A regular HAPS-SMBS node is supposed to have sufficient transmit power levels for signal transmission. Therefore, if harvested energy meets or exceeds the required transmit power, regular HAPS-SMBS utilizes it for signal transmission. However, if the harvested energy is less than the required level, regular HAPS-SMBS supplies a portion of its energy to reach the required transmit power. We will first assume that the regular HAPS-SMBS node has sufficient EH to transmit signals and the second case is described in Section V.

\begin{figure}[]
    \centering
    \includegraphics[width=1\columnwidth]{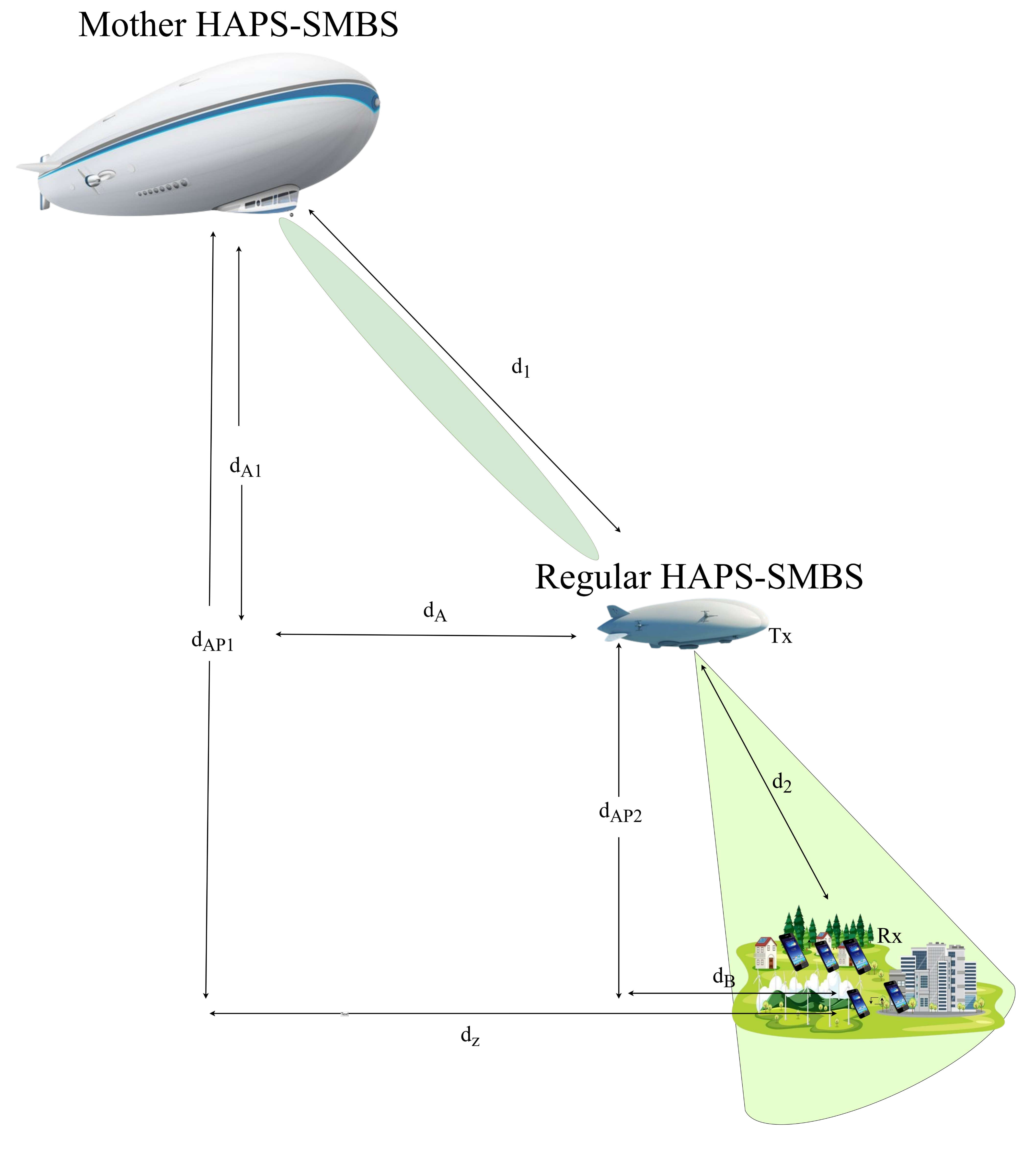}
    \caption{HAPS-SMBS supported by the EH scheme.}
    \label{constellations}
\end{figure}

\subsubsection{Energy Harvesting Model}

In this paradigm, we assume that regular HAPS-SMBS collects energy from mother HAPS-SMBS during the first $\tau T$ time slot, as illustrated in Fig. 3, where $\tau \in [0,1]$ is the EH factor (or it is called the time-switching (TS) factor), $T$ is the transmission block period. This article employs both linear and non-linear EH models. The harvested energy  during the power transfer phase in the linear EH model can be computed by \cite{khennoufa2023wireless,10923658}
\begin{equation}\label{eq:1}
   E_{\text{EH}}^{\text{L}}=\tau \eta P_{\text{t}} |g_{1}|^2 T,  
\end{equation}
where 
\begin{math}\label{eq:1}
g_{1}=\sqrt{G_{\text{t}}G_{\text{r}}} \left( \frac{\lambda}{4 \pi d_{1}} \right)\end{math}, $\eta \in [0,1]$ is the energy conversion efficiency factor, $P_{\text{t}}$ is the transmit power of mother HAPS-SMBS, $d_1$ is the distance between the mother HAPS-SMBS and regular HAPS-SMBS, $\lambda=\frac{c}{f}$ is the wavelength, $c$ is the speed of light in a vacuum, $f$ is the frequency, whereas $G_{\text{t}}$ and $G_{\text{r}}$ are the transmitter and receiver antenna gains, respectively.

\begin{figure}[]
    \centering
    \includegraphics[width=1\columnwidth]{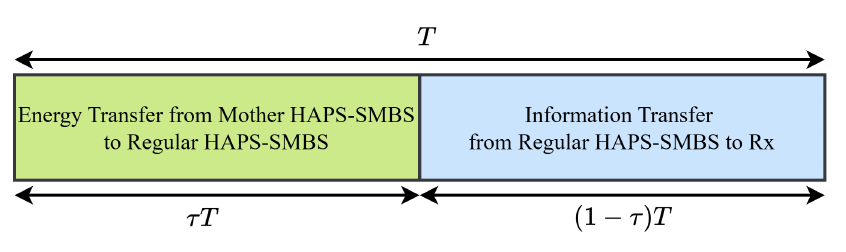}
    \caption{Block diagram of energy harvesting and information transmission for the proposed system.}
    \label{constellations}
\end{figure}

Likewise, to represent the dynamics of the RF energy conversion efficiency for various input power levels, we employ a realistic and parametric non-linear EH model. The logistic function-based non-linear EH model was introduced in \cite{10923658,jiang2019outage,ouamri2023nonlinear} to precisely simulate the non-linearity of practical EH circuits. Hence, the harvested energy for the non-linear EH model can be defined as
\begin{equation}\label{eq:1}
  {\color{black} E_{\text{EH}}^{\text{NL}}=\tau  \Delta  T,  }
\end{equation}
where \begin{math}\label{eq:1} 
\Delta= M \frac{1-e^{-\varsigma P_{\text{t}} |g_{1}|^2}}{1+e^{-\varsigma(P_{\text{t}} |g_{1}|^2- \varrho)}}\end{math}, $M$ is a constant that represents the maximum harvest {\color{black}power} at the EH receiver when the EH circuit is saturated, $\varsigma$ and $\varrho$ are constants determined by the circuit's detailed specifications, such as resistance, capacitance, and diode turn-on voltage.

Subsequently, we obtain the transmit power of regular HAPS-SMBS during $(1 - \tau)T$ time slot (as presented in Fig. 3) for linear and non-linear EH models \hl{as in }\cite{khennoufa2023wireless,10923658,jiang2019outage,ouamri2023nonlinear} \hl{by}
\begin{equation}\label{eq:1}
P_{\text{EH}}^{\text{v}}=\frac{E_{\text{EH}}^{\text{v}}}{(1-\tau)T}, \ \ \text{v}=\{\text{L, NL}\}.
\end{equation}



\subsubsection{Propagation Model}
The regular HAPS-SMBS (transmitter, Tx) communicates with the ground users (receiver, Rx) using the collected energy. Let $x(t)$ represent the transmitted signal by the Tx. The received (noise-free) signal at Rx for linear and non-linear EH models can then be written as in \cite{alfattani2021link} by
\begin{equation}\label{eq:1}
y_{\text{v}}(t)=\alpha_{\text{v}} x(t), \ \ \text{v}=\{\text{L, NL}\},
\end{equation}
where $\alpha_{\text{v}}$ is the wireless channel coefficient, expressed simply with the log-distance path-loss model, which is defined for linear and non-linear EH models by
\begin{equation}\label{eq:1}
\alpha_{\text{v}}=\sqrt{P_{\text{EH}}^{\text{v}}} \ h_{2},
\end{equation}
where \begin{math}\label{eq:1}
h_{2}=\sqrt{G_{\text{t}}G_{\text{r}}} \left( \frac{\lambda}{4 \pi d_{2}} \right)\end{math}, $d_{2}$ is the distance between regular HAPS-SMBS and Rx.

By substituting (1) into (3) and (2) into (3), we obtain the transmit power of regular HAPS-SMBS for linear and non-linear EH models, respectively. Then, we substitute (3) into (5), and we find the wireless channel coefficient $\alpha_{\text{L}}$ for the linear EH model as given by
\begin{equation}\label{eq:1}
   \alpha_{\text{L}}=   \frac{ \sqrt{P_{\text{t}}} \beta G_{\text{t}}G_{\text{r}}}{D} \left( \frac{\lambda}{4  \pi } \right)^2, 
\end{equation}
where \begin{math}\label{eq:1} D=d_{1} d_{2} \end{math}.

Likewise, we obtain the wireless channel coefficient $\alpha_{\text{NL}}$ for the non-linear EH model as given by
\begin{equation}\label{eq:1}
   \alpha_{\text{NL}}=
   \beta \sqrt{ \Delta P_{\text{t}} G_{\text{t}}G_{\text{r}}}  \left( \frac{\lambda}{4 \pi d_{2}} \right), 
\end{equation}
where
{\color{black}
\begin{math}\label{eq:1}\beta=\sqrt{\frac{\tau }{(1-\tau)}} \end{math}.}

Hence, the received power at Rx for the linear EH model can be written as
\begin{equation}\label{eq:1}
   P_r^{\text{L}}=
   P_{\text{t}} \left(\frac{G_{\text{t}}G_{\text{r}} \beta}{ D}\right) ^{2} \left( \frac{\lambda}{4  \pi } \right)^4,  
\end{equation}
and for non-linear EH model can be written as
\begin{equation}\label{eq:1}
   P_{\text{r}}^{\text{NL}}=
   \beta^2  \Delta P_{\text{t}} G_{\text{t}}G_{\text{r}}  \left( \frac{\lambda}{4 \pi d_{2}} \right)^2.  
\end{equation}

To find the signal-to-noise ratio (SNR), we assume that the noise is additive white Gaussian noise (AWGN), such noise mimics the random processes effect in space, where only communication impairment is the white noise \cite{saeed2021point}. Then, the SNR for linear and non-linear EH models can be written as
\begin{equation}\label{eq:1}
    \gamma_{\text{v}}= \frac{P_{\text{r}}^{\text{v}}}{P_{\text{N}}} , \ \ \text{v}=\{\text{L, NL}\}.
\end{equation}

Hence, the data rate at Rx for linear EH model, can be written as 
\begin{equation}\label{eq:1}
   R^{\text{L}}=
  \mathfrak{T} \log_{2} \left( 1 + \frac{P_{\text{t}} \left(\frac{G_{\text{t}}G_{\text{r}} \beta}{ D}\right)^{2} \left( \frac{\lambda}{4  \pi } \right)^4}{P_{\text{N}}} \right),  
\end{equation}
and for non-linear EH model, can be written as 
\begin{equation}\label{eq:1}
   R^{\text{NL}}=
   \mathfrak{T} \log_{2} 
   \left( 1 + \frac{\beta^2 P_{\text{t}} \Delta_1 G_{\text{t}}G_{\text{r}}  \left( \frac{\lambda}{4 \pi d_2} \right)^2}{P_{\text{N}}} \right),  
\end{equation}
where $\mathfrak{T}=B_{\text{W}} (1 - \tau )$, \begin{math}
\Delta_1=M \frac{1-e^{-a \Xi}}{1+e^{-a(\Xi-b)}}
\end{math},
\begin{math}
\Xi=P_{\text{t}} G_{\text{t}}G_{\text{r}} \left( \frac{\lambda}{4 \pi}  \right)^2 \frac{1}{d_1^2}
\end{math}, $P_{\text{N}}=\kappa \Upsilon B_{\text{W}} F$, $\kappa$ is the Boltzmann constant, $\Upsilon$ is the absolute temperature, $B_{\text{W}}$ is the bandwidth, and $F$ is noise figure.


\section{Optimal Positioning for HAPS-SMBS Empowered by EH}
\hl{In this section, the optimal horizontal position of the regular HAPS-SMBS of the proposed architecture is derived. The main purpose of this section is to determine the distance that maximizes the EH and improves the communication performance.
In wireless HAPS-SMBS networks, the horizontal positioning of the regular HAPS-SMBS plays a crucial role in determining both EH and communication performance. While the horizontal distance of a HAPS scenario was derived in} \cite{alfattani2021link}\hl{, the optimization of a two-tier HAPS-SMBS scenario with linear and non-linear EH has not yet been evaluated. 
}
The distance between the energy source (i.e., the mother HAPS-SMBS) and the harvesting node (i.e., regular HAPS-SMBS node) is one of the variables influencing the quantity of wireless harvested energy. By maximizing the harvested energy, the overall system performance, e.g., SNR, and data rate, can be improved. In this section, we derive the optimal distance between the mother HAPS-SMBS and the regular HAPS-SMBS node to maximize the harvested energy and enhance overall system performance. Therefore, in the following Theorems, we obtain the optimal distance for linear and non-linear EH models, respectively.

\textit{\textbf{Theorem 1}}:
We calculate the horizontal distance\footnote{Since the altitude of the aerial platforms is known (e.g., HAPS-SMBS at an altitude around 20 km, and UAV that have a lower altitude (a few hundred meters on the ground)) in the literature as in \cite{alfattani2023multimode,alfattani2021link}, we consider that $d_{\text{A}}$ is unknown in our system. Hence, we need to obtain the optimal distance $d_{\text{A}}$ to reach maximum performance.} $d_{\text{A}}$ to maximize the data rate for the linear EH model. As shown in Fig. 2, given that the total horizontal distance between mother HAPS-SMBS and the regular HAPS-SMBS, and regular HAPS-SMBS and Rx is $d_{\text{z}}=d_{\text{A}}+d_{\text{B}}$. The mother HAPS-SMBS and regular HAPS-SMBS are located at altitudes $d_{\text{AP}1}$ and $d_{\text{AP2}}$, respectively, and horizontally separated by $d_A$ (assuming that all nodes are aligned in the same line with the ($x$ or $y$) axis), i.e., \begin{math} d_{1}=\sqrt{(d_{\text{AP}1}-d_{\text{AP}2})^2+d_{\text{A}}^2} \end{math} and
\begin{math} d_{2}=\sqrt{d_{\text{AP}2}^2+(d_{\text{z}}-d_{\text{A}})^2} \end{math}. Based on the linear EH model, we find $D=d_{1}d_{2}$. Then, the optimal positioning is given by 
\begin{equation}\label{eq:1}
\begin{split}
d_{\text{A}}^{*}= \sqrt[3]{\epsilon_1+\sqrt{\epsilon_1^2+\epsilon_2^3}}+\sqrt[3]{\epsilon_1-\sqrt{\epsilon_1^2+\epsilon_2^3}}-\epsilon_3,
\end{split}
\end{equation}
where \begin{math} \epsilon_1=\frac{-\mathfrak{b}^2}{27\mathfrak{a}^3}+\frac{\mathfrak{b c}}{6\mathfrak{a}^2}- \frac{\mathfrak{d}}{2\mathfrak{a}} \end{math}, \begin{math} \epsilon_2=\frac{\mathfrak{c}}{3\mathfrak{a}}- \frac{\mathfrak{b}^2}{9\mathfrak{a}^2} \end{math}, \begin{math} \epsilon_3=\frac{\mathfrak{b}}{3\mathfrak{a}} \end{math}, \begin{math} \mathfrak{a}=3 \end{math}, \begin{math} \mathfrak{b}=\mathfrak{a} \Lambda_3 \end{math}, \begin{math} \mathfrak{c}=2\Lambda_4 \end{math}, \begin{math} \mathfrak{d}=\Lambda_5 \end{math}, \begin{math} \Lambda_{1}=d_{\text{AP}2}^2 \end{math}, \begin{math} \Lambda_{2}=(d_{\text{AP}1} - d_{\text{AP2}})^2 \end{math}, \begin{math} \Lambda_{3}=-2 d_{\text{z}} \end{math}, \begin{math} \Lambda_{4}=\Lambda_{1}+\Lambda_{2}+d_{\text{z}}^2 \end{math}, \begin{math} \Lambda_{5}=\Lambda_{2} \Lambda_{3} \end{math}, and \begin{math} \Lambda_{6}=\Lambda_{1}\Lambda_{2}+\Lambda_{2} d_{\text{z}}^2 \end{math}. 






\begin{IEEEproof} The equivalent path-loss distance of the linear EH model, denoted by $D$, can be written as 
\begin{equation}\label{eq:1}
\begin{split}
&
D^2=(d_{1} d_{2}) ^2\\& \ \ \ = ((d_{\text{AP1}}-d_{\text{AP2}})^2+d_{\text{A}}^2) (d_{\text{AP2}}^2+(d_{\text{z}}-d_{\text{A}})^2).
\end{split}
\end{equation}
To maximize the data rate, we must minimize $D$ by setting its first derivative to zero as follows\begin{equation}\label{eq:1}
\begin{split}
& \frac{\delta D^2}{\delta d_{\text{A}}}= 
4 d_{\text{A}}^3+ \mathfrak{a} \Lambda_{3} d_{\text{A}}^2+ 2 \Lambda_{4} d_{\text{A}}+ \Lambda_{5}=0.
\end{split}
\end{equation}

By using the Cardano formula \cite{schechter2013cubic}, the real root of the cubic in (15) can be obtained as given in (13).
The second derivative\footnote{In mathematical contexts, we should determine whether points represent maximum or minimum values to analyze curvature and perform accurate estimations. The second derivative provides information on the function's changing slope, concavity, inflection points, and optimization results. } is computed by
\begin{equation}\label{eq:1}
\begin{split}
\frac{\delta^2 D^2}{\delta d_{\text{A}}^2}=  12 d_{\text{A}}^2+ 6 \Lambda_{3} d_{\text{A}}+ 2 \Lambda_{4} >0,
\end{split}
\end{equation}
in which we substitute the roots of (13), revealing that (16) is positive. This $d_{\text{A}}$ value indicates the optimal positioning of regular HAPS-SMBS for achieving the maximum data rate at Rx. The proof is completed.
\end{IEEEproof}

\textit{\textbf{Theorem 2}}:
We calculate the optimal value of $d_{A}$ to maximize the data rate for the non-linear EH model. 
Based on the non-linear EH model in (12), we can observe that $d_{1}$ and $d_{2}$ are separated, unlike the linear model. 
This structural difference prevents the derivation of a closed-form solution, as in the linear case. Therefore, we adopt an approximate approach by solving each component of $d_{1}$ and $d_{2}$ separately and then combining their solutions.
Since both $d_{1}$ and $d_{2}$ are content of $d_{\text{A}}$, the optimal $d_{\text{A}}$ can be computed independently for each. Then, the optimal $d_{\text{A}}$ for the non-linear EH model based on $d_{1}$ is given as
\begin{equation}\label{eq:1}
d_{\text{A}}^{*}= 0.
\end{equation}

Likewise, the optimal $d_{\text{A}}$ for the non-linear EH model based on $d_{2}$ is given as
\begin{equation}\label{eq:1}
d_{\text{A}}^{*}= d_{\text{z}}.
\end{equation}


\begin{IEEEproof} The equivalent path-loss distance of the non-linear EH model, denoted by $d_1$, can be calculated~as 
\begin{equation}\label{eq:1}
\begin{split}
d_{1}^2=(d_{\text{AP1}}-d_{\text{AP2}})^2+d_{\text{A}}^2.
\end{split}
\end{equation}

To maximize the data rate, we must minimize $d_1$ by nulling its first derivative as follows
\begin{equation}\label{eq:1}
\begin{split}
\frac{\delta d_{1}^2}{\delta d_{\text{A}}}= 2d_{\text{A}} =0.
\end{split}
\end{equation}
In this case, the optimal value of $d_{\text{A}}$ based on $d_{1}$ is obtained as follows:
\begin{equation}\label{eq:1}
\begin{split}
d_{\text{A}} =  0.
\end{split}
\end{equation}
The second derivative is given by
\begin{equation}\label{eq:1}
\begin{split}
\frac{\delta^2 d_{1}^2}{\delta d_{\text{A}}^2}= 2 >0.
\end{split}
\end{equation}

Likewise, the equivalent path-loss distance of the non-linear EH model, denoted by $d_{2}$, can be calculated as 
\begin{equation}\label{eq:1}
\begin{split}
d_{2}^2=\Lambda_{1}+(d_{\text{z}}-d_{\text{A}})^2.
\end{split}
\end{equation}

To maximize the data rate, we must minimize $d_{2}$ by nulling its first derivative as 
\begin{equation}\label{eq:1}
\begin{split}
\frac{\delta d_{2}^2}{\delta d_{\text{A}}}= 2d_{\text{A}} -2d_{\text{z}} =0.
\end{split}
\end{equation}
Hence, the optimal value of $d_{\text{A}}$ based on $d_{2}$ is obtained as
\begin{equation}\label{eq:1}
\begin{split}
d_{\text{A}} =  d_{\text{z}}.
\end{split}
\end{equation}
The second derivative of (23) is given by
\begin{equation}\label{eq:1}
\begin{split}
\frac{\delta^2 d_{2}^2}{\delta d_{\text{A}}^2}= 2 > 0.
\end{split}
\end{equation}

 
We find that equations (22) and (26) are positive. The $d_{\text{A}}$ values indicate the optimal regular HAPS-SMBS positioning for achieving the maximum data rate at Rx for the non-linear EH model. The proof is completed.
\end{IEEEproof}

\textcolor{black}{
\textit{Note:} In the non-linear EH model, substituting the solution into $d_{1}$ and $d_{2}$, we obtain \begin{math} d_{1}=(d_{\text{AP}1}-d_{\text{AP}2}) \end{math} and
\begin{math} d_{2}=d_{\text{AP}} \end{math}. This configuration implies that the regular HAPS is positioned at its maximum altitude, i.e., at the closest possible distance to the mother HAPS. As a result, $d_{1}$ is minimized, leading to an improved channel condition and increased harvested energy.}

\section{\hl{JOINT Optimization of HAPS-SMBS Positioning and EH FACTOR}}



\hl{In the proposed EH-enabled two-tier HAPS-SMBS system, the data rate depends on both the amount of harvested energy and the quality of the propagation links. As derived in Sections II and III, the data rate is governed by two key design parameters: the EH factor $\tau$, representing the portion of the transmission block allocated to wireless EH, and the horizontal positioning distance $d_{\text{A}}$ between the mother HAPS-SMBS and the regular HAPS-SMBS, which directly affects both the harvested energy and signal attenuation. While only the HAPS position is considered in }\cite{alfattani2021link}\hl{, the study does not address a two-tier system with EH nor the joint optimization of $\tau$ and $d_{\text{A}}$. 

In contrast, practical deployment scenarios require both the HAPS-SMBS position and the EH factor to be optimized simultaneously to fully exploit the benefits of EH-enabled HAPS-SMBS, since a fundamental trade-off exists between them that governs the amount of harvested energy and the system's performance. Therefore, the objective of this work is to jointly optimize the EH factor ($\tau$) and the regular HAPS-SMBS positioning ($d_{\text{A}}$) to maximize the data rate. The optimization is performed under practical system constraints, including EH duration, HAPS-SMBS distances, and transmit power, capturing the trade-off between EH and information transmission.
}


\subsection{Problem Formulation}

\hl{Based on the objective defined above and the data rate expressions in Section II, the joint optimization problem of $\tau$ and $d_{\text{A}}$ is formulated as }
\begin{subequations}
\begin{align}
(\text{P}): & \operatorname*{max}_{d_{\text{A}}, \tau} R^{\text{v}}, \ \ \text{v}=\{\text{L, NL}\} ,
\\& \quad 
\text{s.t.:} \ 0 < \tau < 1,
  \\&  \ \ \ \ \ \ \ d_{\text{A}} > 0,
  \\& \ \ \ \ \ \ \ d_{\text{AP1}} > d_{\text{AP2}}, 
  \\& \ \ \ \ \ \ \ d_{\text{AP1}} > d_{\text{z}},   
    \\& \ \ \ \ \ \ \ P_{\text{t}} \geq P_{\text{t,min}},
 \end{align}
\end{subequations}    
where $P_{\text{t,min}}$ is the minimum required to transmit power, $ R^{\text{v}}$ is selected as the optimization function and is driven by our objective to maximize the system performance, and $\tau$ is the permissible range value i.e., $\tau \in [0, 1]$. Moreover, $d_{\text{A}}>0$ must be exactly greater than zero to maintain the physical feasibility of structuring usable communication and power transmission networks between platforms. The altitude of mother HAPS-SMBS should be higher than regular HAPS-SMBS, i.e., $d_{\text{AP1}} > d_{\text{AP2}}$. To maintain operational constraints, the constraint $P_{\text{t}} \geq P_{\text{t},\text{min}}$ \hl{ensures that the transmit power remains above the minimum operational threshold.} The objective function maximizes the data rate based on the EH factor ($\tau$) and the distance ($d_{\text{A}}$). 
Although the conditions in the problem are linear, it is worth noting that this problem is a nonlinear programming problem (NLPP) as well as a non-convex optimization problem due to the nature of the objective function, which contains complex ratios depending on the numerous variables, such as $d_{\text{A}}$ and $\tau$.
This non-convexity arises from nonlinear relationships introduced by logarithmic and exponential terms; these introduce complicated dependencies between variables that eventually introduce multiple local optima.
This problem is difficult to solve using mathematical techniques because of its non-convex nature. 

\subsection{\hl{Proposed Solution Approaches}}
\hl{To solve the above joint optimization problem, we consider two complementary solution approaches. The first approach decomposes the problem into two sub-problems, which are solved separately using an IDFA. This iterative approach is selected for its ease of execution and usefulness in dealing with low-dimensional non-convex problems. Although suboptimal, it converges quickly and achieves near-optimal performance with low computational complexity. However, based on the previous literature}~\cite{zhong2022deep}\hl{, splitting the problem may sometimes not provide an exact solution. Therefore, we propose a second approach that leverages Q-learning to adaptively optimize the variables jointly and efficiently handle dynamic environments. In the following subsections, we explain the two proposed solutions in detail.}




\subsubsection{\hl{Proposed Solution 1: Iterative Distance and EH Factor Algorithm} }
\hl{In this subsection}, we solve the problem (P) using the coordinate descent approach~\cite{razaviyayn2013unified}, and we split the optimization problem into two subproblems. 
\hl{In the first subproblem, we solve the problem (P) using IDFA for the case where $d_{\text{A}}$ is given, and in the second optimization subproblem, $\tau$ is assumed to be given.}

\textit{\textbf{Proposition 1}}: If we fix the $\tau$ in (P), we will have the $d_{\text{A}}$ sub-problem as follows:
\begin{subequations}
\begin{align}
(\text{P1}): & \operatorname*{max}_{d_{\text{A}}} R^{\text{v}}, \ \ \text{v}=\{\text{L, NL}\} ,
\\& \quad 
\text{s.t.:} \ \text{(27c), (27d), (27e), (27f)}.
 \end{align}
\end{subequations}

Likewise, for a given $d_{\text{A}}$ in (P), we will have the following $\tau$ sub-problem:
\begin{subequations}
\begin{align}
(\text{P2}): & \operatorname*{max}_{\tau} R^{\text{v}}, \ \ \text{v}=\{\text{L, NL}\} ,
\\& \quad 
\text{s.t.:} \ \text{(27b), (27f)}.
 \end{align}
\end{subequations}  

To find optimal solutions for $d_{\text{A}}$ and $\tau$ and address the optimization problem, we use an iterative optimization method using the coordinate descent approach. The problem is solved by optimizing two subproblems alternately: first, optimizing $d_{\text{A}}$ while keeping $\tau$ constant, and then optimizing $\tau$ given $d_{\text{A}}$.

To begin, we set up $\tau$ and $d_{\text{A}}$. At every iteration, $d_{\text{A}}$ is optimized with respect to the data rate ($R^{\text{v}}$) within the range of $d_{\text{A}}$ values. Next, with the optimized $d_{\text{A}}$, we proceed to optimize $\tau$ by calculating the $R^{\text{v}}$ over the range of $\tau$ values. This is repeated using these optimal values as the starting point of the next iteration until convergence. 
\textcolor{black}{
The IDFA runs in $\mathcal{O}(i \times j)$ time per iteration, where $i$ and $j$ are the number of grid points for $d_{\text{A}}$ and $\tau$, respectively. 
}
The associated algorithm is summarized in Algorithm 1.

\begin{algorithm}[]
\caption{Iterative Distance and EH Factor Algorithm (IDFA).}
\begin{algorithmic}[1]

\State \textbf{Input:} $P_{\text{t}}$, $G_{\text{t}}$, $G_{\text{r}}$, $f$, $\eta$, $d_{\text{AP1}}$, $d_{\text{AP2}}$, $d_{\text{z}}$, $B_{\text{w}}$, $F$, maximum number of iterations $N_{\text{max}}$, and tolerance $\varepsilon > 0$.

\State \textbf{Initialize:} Set initial values $\tau^{(0)}$ and $d_{\text{A}}^{(0)}$, and previous data rate $R^{\text{v}}_{\text{prev}} \gets 0$.

\For{ $N_{\text{max}}$}

    \State \textbf{Optimize $d_{\text{A}}$ (fix $\tau$):}
    \State Define $d_{\text{A}}$ range and constraints as in (27c)--(27f).
    
    \For{each $d_{\text{A}}$ in range}
        \State Compute distances: $d_1$ and $d_2$
        \State Compute $R^{\text{v}}$ using equations (11) and (12)
    \EndFor
    
    \State Update $d_{\text{A}}$ to maximize $R^{\text{v}}$ based on (11) and (12)

    \State \textbf{Optimize $\tau$ (fix $d_{\text{A}}$):}
    \State Define $\tau$ range and constraints as in (27b) and (27f)
    
    \For{each $\tau$ in range}
        \State Compute $R^{\text{v}}$ using equations (11) and (12)
    \EndFor
    
    \State Update $\tau$ to maximize $R^{\text{v}}$ based on (11) and (12)

    \If{$R^{\text{v}} - R^{\text{v}}_{\text{prev}} < \varepsilon$}
        \State \textbf{break}
    \EndIf
    
    \State $R^{\text{v}}_{\text{prev}} \gets R^{\text{v}}$

\EndFor

\State \textbf{Output:} Optimal $d_{\text{A}}$, $\tau$, and maximum $R^{\text{v}}$.

\end{algorithmic}
\end{algorithm}





\subsubsection{\hl{Proposed Solution 2: $Q$-Learning Design}}
\label{subsubsec:QLearning}

Although the problem (P) can be easily solved by dividing it into two sub-problems and providing a sub-optimal solution, we want to verify the results through joint optimization without resorting to dividing the problem into two subproblems. Therefore, we propose $Q$-learning to solve the problem and verify the effectiveness of the IDFA approach. Therefore, we propose employing a reinforcement learning approach, a set of machine learning algorithms, to determine the optimal values for $\tau$ and $d_{\text{A}}$ jointly, and verify the effectiveness of the IDFA approach.
Reinforcement learning is distinguished by a risk and reward framework in which the agent gathers information from the environment, makes decisions, and receives rewards or penalties based on the correctness of its actions. The previous study, as described in \cite{abubakar2019q}, demonstrated that reinforcement learning is useful in controlling decision-making processes with a large number of options. This is the reason behind its application in this situation.

In this paper, we employ the $Q$-learning algorithm as described in \cite{abubakar2019q}. $Q$-learning is one of the most widely used reinforcement learning algorithms and has proven effective in adapting to dynamic environments. The fundamental framework of $Q$-learning comprises six key components: (i) agent, (ii) environment, (iii) action, (iv) state, (v) reward/penalty, and (vi) action-value table. The agent, acting within a specific environment, takes actions intending to maximize rewards or minimize penalties. The state that results from each action is evaluated, along with any related reward or penalty. Next, updates are made to the action-value table, a crucial component that houses rewards and penalties for every possible action and state, in accordance with following this rule
\begin{equation}\label{eq:1}
\begin{split}
& Q(s_t, a_t) :=    Q(s_t, a_t) \\& \ \ \ \ \ \ \ \ \ \ \  \
+ \beth \left[ \zeta_{t}^{\text{v}} + \phi \max_{a} Q(s_{t+1}, \upsilon) - Q(s_t, a_t) \right],
\end{split}
\end{equation}
where $s_{t}$ and $s_{t+1}$ are the current and next states, respectively, $\zeta_{t}^{\text{v}}$ is the received reward at time step $t$ for linear and non-linear EH models, $\beth$ is a learning rate, which is 0 $\leq \beth \leq$ 1, $\phi$ is a discount factor, which is 0 $\leq \phi \leq$ 1, $\upsilon$ is the set of all possible actions, $a_t$ is the taken action.


In order to solve our problem with $Q$-learning, we discretize the continuous decision/optimization variables, such that $d_{\text{A}}$=[$d_{\text{A1}}$, $d_{\text{A2}}$,...,$d_{\text{A}\mathfrak{i}}$] and $\tau$=[$\tau_1$, $\tau_2$, ..., $\tau_{\mathfrak{j}}$], where $\mathfrak{i}$ and $\mathfrak{j}$ are the numbers of elements included in the vectors of $d_{\text{A}}$ and $\tau$, respectively, somehow indicating the resolution of the discretization. 
The design of the algorithm is given in the following paragraphs.

\begin{itemize}
    \item \text{\textbf{States}}: The state matrix of the proposed $Q$-learning algorithm is an $\mathfrak{i}$ and $\mathfrak{j}$ matrix and composed of 2-tuples of all the one-to-one mapping between $d_{\text{A}}$ and $\tau$ such that 
\begin{equation}
  S=[d_{\text{A}\mathfrak{i}}, \tau_{\mathfrak{j}}], \ \ \forall_{\mathfrak{i},\mathfrak{j}}. 
\end{equation}

There can be alternative designs as well since the state design in (31) is exhaustive and prone to result in infeasible computational costs.

\item \text{\textbf{Actions}}: Similar to the states, the action matrix of the proposed $Q$-learning algorithm can be designed as  
\begin{equation}
  A=[d_{\text{A}\mathfrak{i}}, \tau_{\mathfrak{j}}], \ \ \forall_{\mathfrak{i},\mathfrak{j}}. 
\end{equation}

The basic principle/relation between states and actions is that any taken action should be capable of changing the state, and which is fully satisfied with this set of state-action designs.

\item \text{\textbf{Reward Function}}:
Since this is an optimization problem and the average data rate $R^{\text{v}}$ for linear and non-linear EH models, are tried to be optimized/maximized, it would be wise to select this function as the reward function of the proposed $Q$-learning algorithm, as given by
\begin{equation}\label{eq:1}
   \zeta_{t}^{\text{L}} =
  \mathfrak{T} \log_{2} \left( 1 + \frac{P_{\text{t}} \left(\frac{G_{\text{t}}G_{\text{r}} \beta}{ D}\right)^{2} \left( \frac{\lambda}{4  \pi } \right)^4}{P_{\text{N}}} \right),  
\end{equation}
and
\begin{equation}\label{eq:1}
   \zeta_{t}^{\text{NL}} =
   \mathfrak{T} \log_{2} 
   \left( 1 + \frac{\beta^2 P_{\text{t}} \Delta_1 G_{\text{t}}G_{\text{r}}  \left( \frac{\lambda}{4 \pi d_2} \right)^2}{P_{\text{N}}} \right). 
\end{equation}


\item \text{\textbf{Agent}}:
Considering the set of states and actions, the agent for this particular $Q$-learning design is selected to be regular HAPS-SMBS, since it has full control over the decision/optimization variables. Therefore, regular HAPS-SMBS chooses the 2-tuples of [d$_{\text{A}\mathfrak{i}}$, $\tau_\mathfrak{j}$] from (32) and calculates the corresponding reward, through (33) and (34) followed by determining the states via (31). 

\end{itemize}



\section{Flight Mission Optimization: Transmit Power and Energy Harvesting}
\hl{In this section, we focus on ensuring reliable communication of the regular HAPS-SMBS while preserving its limited onboard energy, which is critical for sustaining flight operations. 
First, we maximize the transmit power of the regular HAPS-SMBS by using the minimum required energy from its inventory. This approach guarantees that the HAPS-SMBS can maintain the required transmission power despite limited harvested energy, while minimizing the impact on its onboard reserves that are primarily needed for flight. 
Second, we calculate the amount of EH during the flight mission. This allows us to quantify the contribution of harvested energy to the transmission process and to evaluate how effectively the regular HAPS-SMBS can rely on EH instead of depleting stored energy. Such analysis also enables estimating the potential extension of operational lifetime for the HAPS-SMBS, highlighting the trade-off between transmission and flight energy consumption.
}

\subsection{Maximizing Transmit Power}
In scenarios where the collected energy by regular HAPS-SMBS is insufficient to transmit signals to Rx, we propose borrowing energy from the regular HAPS-SMBS reserves to meet the required transmission power levels. Accordingly, the new transmit power is expressed by
\begin{equation}
    P_{\text{a}}^{\text{v}} = \frac{E_{\text{a}}^{\text{v}} + E_{\text{EH}}^{\text{v}}}{(1-\tau)T} , \ \ \text{v}=\{\text{L, NL}\},
\end{equation}
where $E_{\text{a}}^\text{v}$ is the energy utilized from the regular HAPS-SMBS inventory in joule. 
Since $E_{\text{a}}^{\text{v}}$ and $ E_{\text{EH}}^{\text{v}}$ are in joule and they divided by $T$ as in (35), so, $P_{\text{a}}^{\text{v}}$ should be in Joule/s or it can be obtained in Watt or dBm. 

Consequently, it becomes imperative to draw upon a minimal amount of power from the regular HAPS-SMBS inventory to maintain the maximum energy of the regular HAPS-SMBS. 
By considering all parameters of both linear and non-linear EH to be fixed, the amount of harvested energy changes according to the transmit power of mother HAPS-SMBS. Thus, the minimum energy used from the regular HAPS-SMBS inventory can be formulated as
\begin{subequations}
\begin{align}
& \operatorname*{min}_{E_{\text{EH}}^{\text{v}}} E_{\text{a}}^{\text{v}}  ,
\ \ \text{v}=\{\text{L, NL}\}, 
&\\  \text{s.t.:} 
& \ \ \ P_{\text{a}}^{\text{v}} \geq P_{\text{Req}},
 \end{align}
\end{subequations}
where $P_{\text{Req}}$ is the required transmit power.

\textit{\textbf{Proposition 2}}: The minimum value of $E_{\text{a}}^{\text{v}}$ that meets the transmission requirement can be determined as follows.
\begin{equation}
   E_{\text{a}}^{\text{v}} = \operatorname*{max} \{(1-\tau)T P_{\text{Req}} - E_{\text{EH}}^{\text{v}},0\},
\end{equation}
which implies that the regular HAPS-SMBS will stop using its energy if the EH exceeds the power required.

\begin{IEEEproof} The used energy from the regular HAPS-SMBS inventory is based on the quantity of the harvested energy, which is changed according to the transmit power of the mother HAPS-SMBS.
Hence, the problem of (36) can be solved based on the following calculation. By substituting (35) into (36b) and after some calculation and simplification, we obtain
\begin{equation}
   E_{\text{a}}^{\text{v}} \geq (1-\tau)T P_{\text{Req}} - E_{\text{EH}}^{\text{v}}.
\end{equation}

Consequently, the minimum value of $E_{\text{a}}^{\text{v}}$ that complies with the transmission requirement can be given as
\begin{equation}
   E_{\text{a}}^{\text{v}} = \operatorname*{max} \{(1-\tau)T P_{\text{Req}} - E_{\text{EH}}^{\text{v}},0\}.
\end{equation}
The proof is completed.
\end{IEEEproof}

\subsection{Amount of Harvested Energy During the Flight Mission }
As referenced in previous research in \cite{gangula2018landing}, the power consumption of aerial platforms during flight is considerably higher than when they are acting as a base station or relay. This reveals a crucial trade-off between flight and transmission power, suggesting that conserving flight power can extend the lifespan of aerial platforms in wireless networks. According to the literature, HAPS-SMBS can stay airborne for a few days or months (e.g., Stratobus). Since regular HAPS-SMBS can harvest energy from mother HAPS-SMBS, we need to calculate the amount of harvested energy during its flight mission and estimate how much the use of this energy during its flight in the air. This enables us to estimate the extra flight time for HAPS-SMBS when they use EH and save their energy for flight. Hence, the total amount of energy collected during the flight mission can be expressed as
\begin{equation}\label{eq:1}
\begin{split}
& E_{\text{Total}}=  \frac{E_{\text{EH}}^{\text{v}} \times T_{\text{T,v}}}{\tau T},
\end{split}
\end{equation}
where $T_{\text{T,v}}$ is the total flight time and $\tau T$ is the energy transfer time.

 \section{Numerical Results} 
\hl{In this section, the performance of the proposed optimization frameworks and system design is evaluated through numerical simulations. Specifically, we evaluate the effectiveness of the optimal positioning strategy derived from the theoretical analysis (i.e., Theorems 1 and 2), the joint optimization using IDFA and $Q$-learning, and the transmit power maximization and amount of EH on system performance. The evaluation focuses on key performance indicators, including harvested energy and data rate under varying operational parameters such as distance, carrier frequency, flight duration, and antenna gain.}
Unless otherwise specified, the simulation parameters are set as $d_{\text{AP1}}=22$~km, $d_{\text{AP2}}=18$~km, $d_{\text{z}}=20$~km, $G_{\text{t}}=43.2$~dBi, $G_{\text{r}}=40$~dBi, $f=2.45$~GHz, $\eta=0.95$, $\tau=0.1$, $T=1$~s, $K=1.38\times10^{-23}$~J.K$^{-1}$, $\Upsilon=300$~K, $B_{\text{W}}=800$~MHz, and $F=7$~dB~\cite{alfattani2021link}.

\subsection{\hl{Performance Evaluation of Proposed Optimization }}
\hl{In this subsection, we evaluate the effectiveness of the proposed optimization frameworks, including the optimal positioning of the regular HAPS-SMBS ($d_{\text{A}}$) derived in Theorems 1 and 2, as well as the joint optimization of positioning and the EH factor using the IDFA algorithm (Algorithm~1) and the $Q$-learning approach described in Subsubsection~IV-B-2. The main objective is to quantify the performance gains and compare them with baseline schemes, such as without EH and random-selection strategies.}

\begin{figure}[]
    \centering
    \includegraphics[width=1\columnwidth]{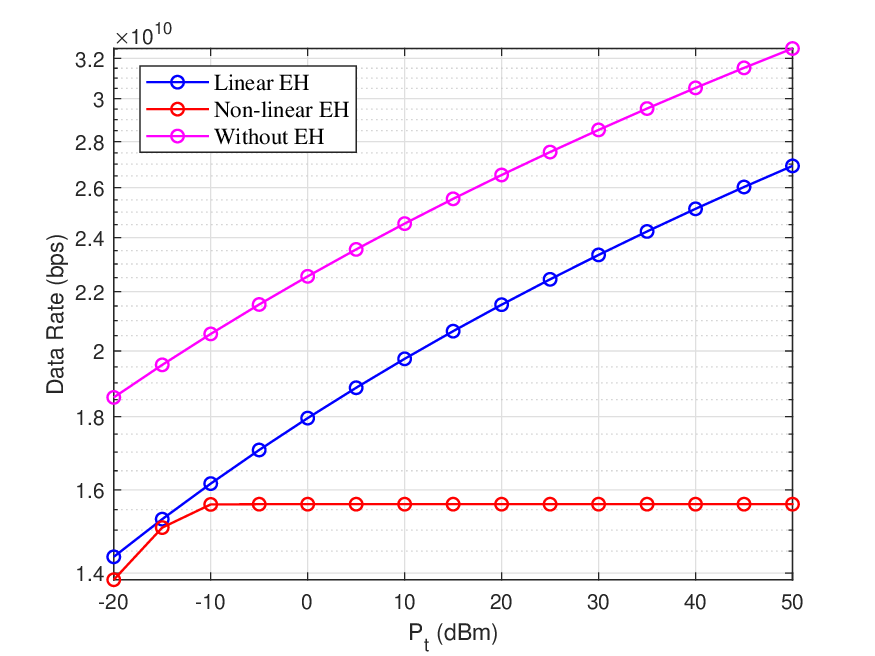}
    \caption{Data rate w.r.t. transmit power ($P_{\text{t}}$): Comparison between EH and without EH.}
    \label{constellations}
\end{figure}

\begin{figure}[]
    \centering
    \includegraphics[width=1\columnwidth]{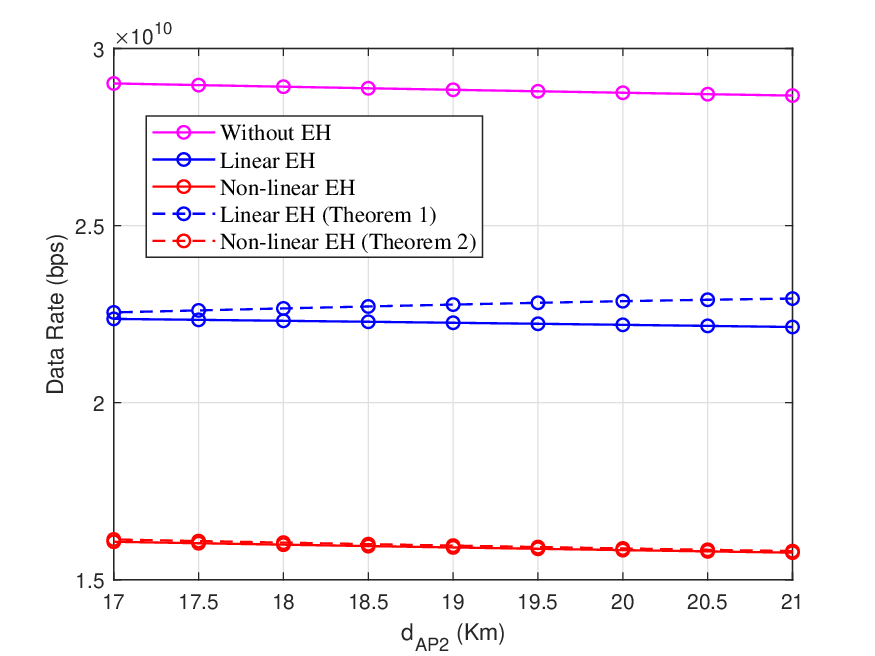}
    \caption{Data rate w.r.t. $d_{\text{AP2}}$ (Km): Comparison between the EH with and without optimization (Theorems 1 and 2) and without EH.}
    \label{constellations}
\end{figure}

Fig. 4 \hl{compares the data rates of linear and non-linear EH models with a scenario without EH}. For fairness, the transmitted power in the system without EH is set equal to the total power $\textit{P}_{\text{t}}$, while in the system with EH, it is $\sfrac{\textit{P}_{\text{t}}}{\tau}$, as mentioned in \cite{khennoufa2023wireless} and \cite{kara2021eror}. It can be seen that the non-linear model achieves lower gains and saturates at higher transmit power levels. This is different from the linear model, which does not experience the same limitations because it considers ideal conditions. This reflects practical experiments showing that EH circuits have non-linear input and output characteristics. \hl{Therefore, the linear EH assumption may not adequately represent real-world scenarios. The figure also shows that the data rate without EH is higher than that of EH schemes; however, the EH models still provide satisfactory performance. To further enhance efficiency, optimization techniques are essential. This underscores the importance of incorporating such techniques to improve overall system performance.}

For this reason, in Fig. 5, we obtain the data rate performance based on the optimal regular HAPS-SMBS positioning as given in Theorems 1 and 2. This figure present the data rate w.r.t. $d_{\text{AP2}}$ (Km). We observe that systems without EH still outperform EH systems. Nevertheless, Theorems 1 and 2 have an improvement rate of $0.8$ to $3.6\%$ and $0.25$ to $0.4\%$ for linear and non-linear EH models, respectively, compared to those without optimization. \hl{These results highlight the importance of HAPS-SMBS positioning in enhancing the amount of EH and overall system performance. Furthermore, it confirms the effectiveness of the optimal HAPS-SMBS positioning strategy derived in Theorems~1 and~2.}

\hl{Therefore}, IDFA and $Q$-learning are good options for determining the optimal positioning for regular HAPS-SMBS to improve the amount of EH and data rate. Another important factor that impacts the amount of EH is the EH factor. To obtain the highest data rate performance, we employ IDFA and $Q$-learning for the joint optimization of $d_{\text{A}}$ and $\tau$, as described in \hl{~IV-B-1 and~IV-B-2}. We compare the $Q$-learning algorithm's convergence speed through the average reward per episode during the learning process in Fig. 6 when $P_{\text{t}}=30$ dBm. \hl{We observe that the average reward quickly increased from a low initial level and stabilized after several hundred episodes, indicating that the proposed $Q$-learning–based optimization converges efficiently and achieves stable performance throughout training.}

\begin{figure}[]
    \centering
    \includegraphics[width=1\columnwidth]{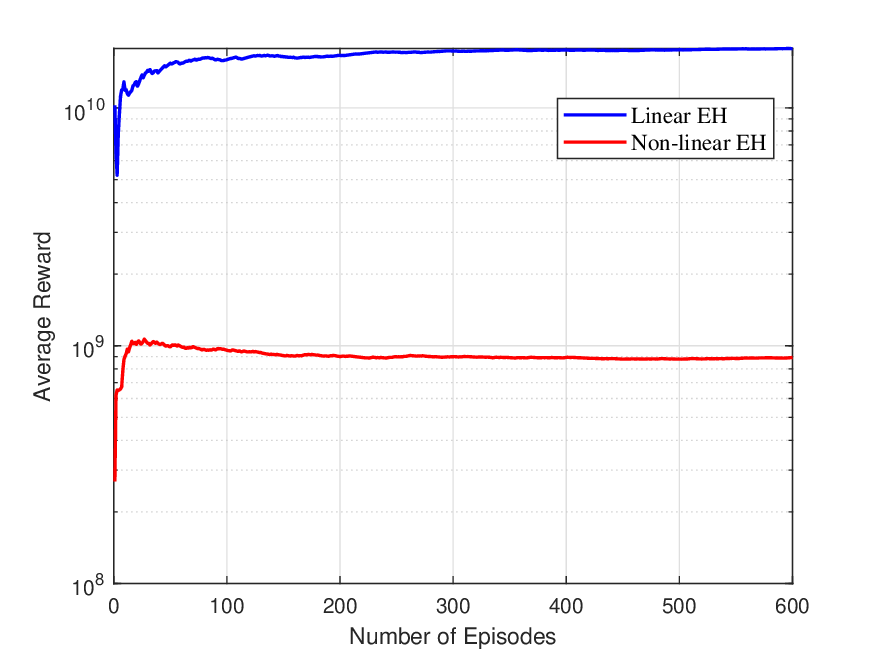}
    \caption{Average reward w.r.t. number of episodes: Comparison between linear and non-linear EH.}
    \label{constellations}
\end{figure}

\begin{table}[]
    \centering
    \caption{\(d_A\) and \(\tau\) values in simulation scenarios.}
    \begin{tabular}{|c|c|c|}
        \hline
        \textbf{Scenario} & \( d_A \) & \( \tau \) \\
        \hline
        \multirow{3}{*}{Scenario A} & \multirow{3}{*}{8 Km} & 0.1 \\ \cline{3-3}
         &  & 0.4 \\ \cline{3-3}
         &  & 0.9 \\ 
        \hline
        \multirow{3}{*}{Scenario B} & \multirow{3}{*}{10 Km} & 0.1 \\ \cline{3-3}
         &  & 0.4 \\ \cline{3-3}
         &  & 0.9 \\ 
        \hline
        \multirow{3}{*}{Scenario C} & \multirow{3}{*}{15 Km} & 0.1 \\ \cline{3-3}
         &  & 0.4 \\ \cline{3-3}
         &  & 0.9 \\ 
        \hline
    \end{tabular}
\end{table}

\begin{figure}[]
\centering
    \centering
    \includegraphics[width=1\columnwidth]{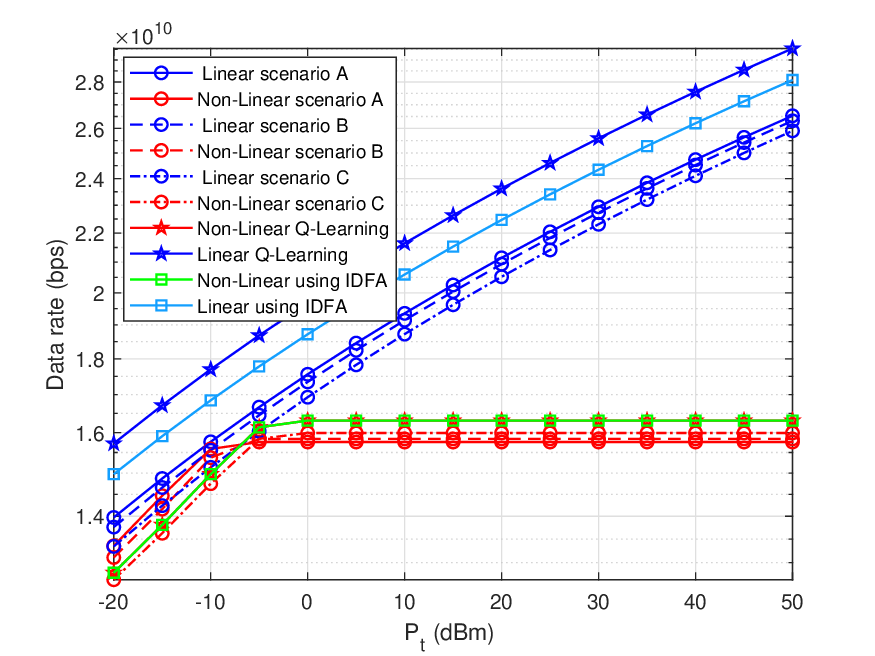}
    \caption{Data rate comparison between linear and non-linear EH with and without $Q$-learning and IDFA.}
    \label{constellations}
\end{figure}

\begin{figure}[]
    \centering
    \includegraphics[width=1\columnwidth]{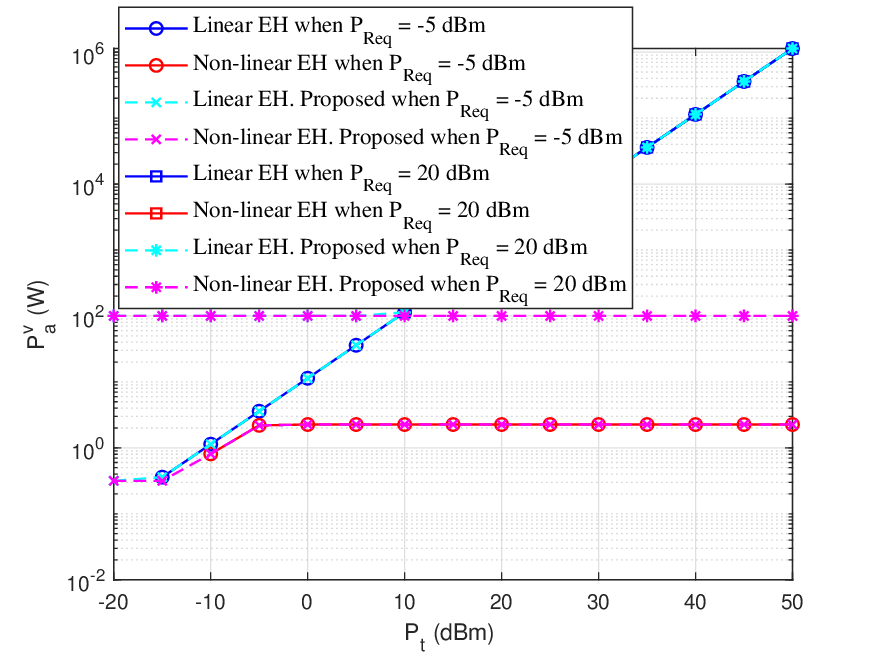}
    \caption{Transmit power comparison between linear and non-linear EH with and without power optimization.}
    \label{constellations}
\end{figure}

Accordingly, in Fig. 7, we compare $Q$-learning with joint optimization $d_{\text{A}}$ and $\tau$ to those without $Q$-learning (i.e., a random selection of $d_{\text{A}}$ and $\tau$ as shown in the Table. 1)  and \hl{IDFA}, when $P_{\text{t}}=30$ dBm. Random selection depends on fixing one of the values, such as $d_{\text{A}}$, changing the  $\tau$ value randomly, and then trying to find the optimal performance among all the values. The process was repeated with random selections for different distances $d_{\text{A}}$ as shown in Table 1.
\hl{The figure shows that the random selection strategy does not yield significant performance gains. In contrast, $Q$-learning and IDFA achieve higher improvements, with rates of $7$ to $11\%$ and $5.5$ to $7\%$ for the linear EH model using $Q$-learning and IDFA, respectively, and $0.8$ to $1.8\%$ for the non-linear EH model. It is observed that $Q$-learning and IDFA perform identically in the non-linear model, while a small gap exists in the linear model, indicating that $Q$-learning may provide a better solution in some cases.}
Even though this improvement might seem modest, \hl{this framework provides a valuable approach}, reduces energy consumption, and can be improved in future work. \hl{Overall, these results confirm the effectiveness of the joint optimization framework based on IDFA and $Q$-learning in enhancing EH system performance, particularly for the linear model using $Q$-learning.}

According to the above findings, the EH performance is improved by rates ranging from $0.8$ to $3.6\%$ and $0.25$ to $0.4\%$ for linear and non-linear EH models, respectively, using Theorems 1 and 2. $Q$-learning produces improvement by rates ranging from $7$ to $11\%$ and $5.5$ to $7\%$ for the linear EH model with $Q$-learning and IDFA, respectively, and $0.8$ to $1.8\%$ for the non-linear EH model with $Q$-learning and IDFA.
To reach the maximal performance, we propose the use of minimum energy from the regular HAPS-SMBS inventory as given in the problem formulation (36). This issue is solved as given in (37). Consequently, when the amount of EH is insufficient, the regular HAPS-SMBS must add a minimum amount of energy from its inventory. If the harvested energy exceeds the required transmission threshold, the regular HAPS-SMBS will cease using its energy. Hence, the optimal transmit power of (35) w.r.t. transmit power of mother HAPS-SMBS ($P_{\text{t}}$), is presented in Fig. 8 with different required transmit power at regular HAPS-SMBS. 
This figure provides evidence that the conventional EH curves (linear and non-linear) are missing at lower transmit power. 
In contrast, the proposed EH for linear and non-linear systems involves a transmit power saturated at fixed values in both required transmit power ($P_{\text{Req}}=-5$ dBm and $P_{\text{Req}}=20$ dBm) at that levels. This indicates that some power is being used from the regular HAPS-SMBS at these levels.
Subsequently, it is observed that the transmit power of regular HAPS-SMBS ($ P_{\text{a}}^{v}$) increases with the increased transmit power of mother HAPS-SMBS ($P_{\text{t}}$). As the linear EH increases, the non-linear EH reaches saturation at high $P_{\text{t}}$ values. In general, the saturation of the non-linear EH model at high $P_{\text{t}}$ values shows how important it is to comprehend the limitations and characteristics of EH circuits in real-world uses. This figure shows that the transmit power of the harvested energy reaches the required levels. Although this increase achieves significant performance gains, in practical applications, when the harvested energy exceeds the required level, the additional energy can be stored and used for other transmissions or other purposes. \hl{This validates the proposed transmit power optimization strategy and its role in minimizing dependence on stored energy while maintaining reliable communication.}

\begin{figure}[]
    \centering
    \includegraphics[width=1\columnwidth]{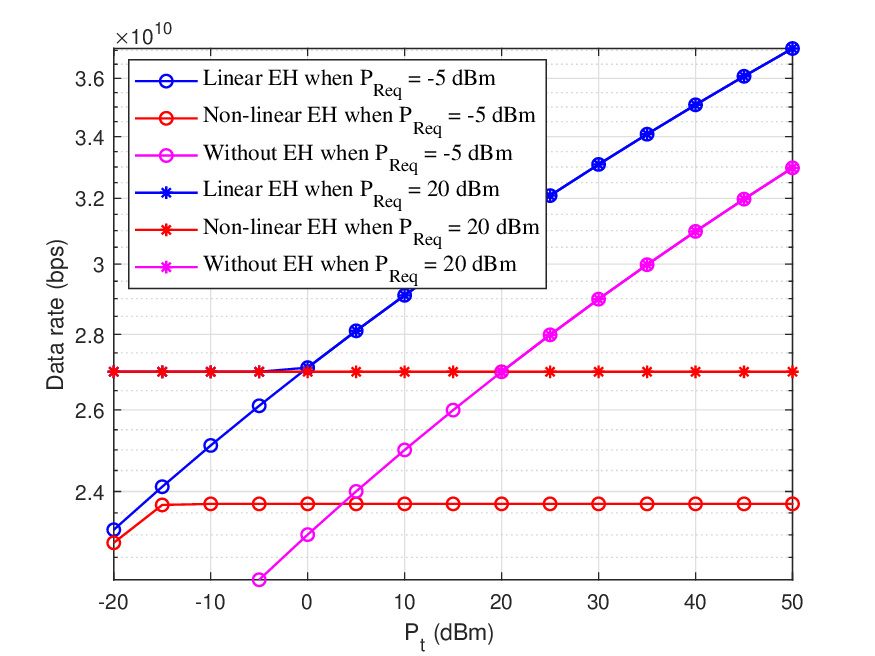}
    \caption{Data rate comparison between linear and non-linear EH with and without power optimization.}
    \label{constellations}
\end{figure}

To provide further clarification, we display the data rate w.r.t. transmit power ($P_{\text{t}}$) with two threshold levels ($P_{\text{Req}}=-5$ dBm and $P_{\text{Req}}=20$ dBm) in Fig. 9. We compare our results to those without the EH system. It can be easily seen that the proposed system outperforms the system without EH in both threshold levels. 

\hl{Overall, the numerical results validate the effectiveness of the proposed optimization frameworks. Specifically, the optimal positioning of the regular HAPS-SMBS based on Theorems 1 and 2 improves data rate performance by $0.8$–$3.6\%$ for linear EH and $0.25$–$0.4\%$ for non-linear EH. Furthermore, joint optimization using IDFA and $Q$-learning provides additional performance gains, with $Q$-learning achieving rates of $7$ to $11\%$ improvement in linear EH scenarios and $0.8$ to $1.8\%$ in non-linear EH scenarios. $Q$-learning outperforms IDFA in the linear EH model, with improvement rates ranging from $1.5$ to $4\%$ under intensive training conditions based on Fig. 7. Finally, the transmit power optimization strategy effectively minimizes dependence on stored energy while maintaining reliable communication. These results collectively demonstrate that the proposed methods significantly enhance EH system performance and data rate compared to conventional schemes.}

\subsection{\hl{Parameter Sensitivity}}
\hl{In contrast to the previous subsection, this subsection investigates system performance sensitivity to key parameters when the proposed optimizations are applied, such as antenna gain, operating frequency, communication distance, and flight time.}

In order to further investigate the EH system in regular HAPS-SMBS, we measured the harvested energy's performance using (1) and (2) with different frequencies, distances, and antenna gains.
In Fig. 10 (a), we compare the output energy for EH with different frequencies. The results show that the increase in the frequency reduces the amount of the EH, where the non-linear model is saturated at lower transmit power values. This can be explained as follows. Higher frequencies experience more attenuation, which limits the harvested energy. Furthermore, higher frequencies may be more vulnerable to environmental factors such as atmospheric absorption and interference, reducing the efficiency of EH systems.
Additionally, we measure the output energy w.r.t. the receiver antenna gain ($G_{\text{r}}$) with different $d_1$ (Km) in Fig. 10 (b), when $P_{\text{t}}=30$ dBm. This figure shows that the non-linear EH model has a negative impact compared to the linear model because there is saturation at lower $G_{\text{r}}$ values. This shows the impact of EH circuits in the real world. It is also observed that an increase in receiver antenna gain increases the amount of EH, whereas increasing $d_1$ reduces it. These results indicate the necessity of careful selection of the positioning of the regular HAPS-SMBS.

\begin{figure}[]
\centering
\subfloat[]{%
\includegraphics[width=1\columnwidth]{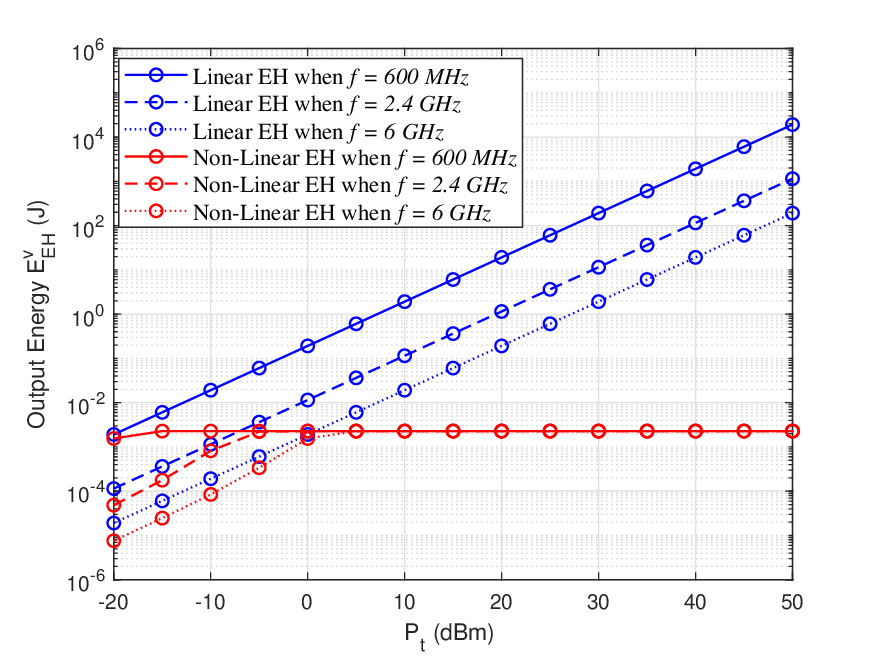}
}
\hfill
\subfloat[]{%
\includegraphics[width=1\columnwidth]{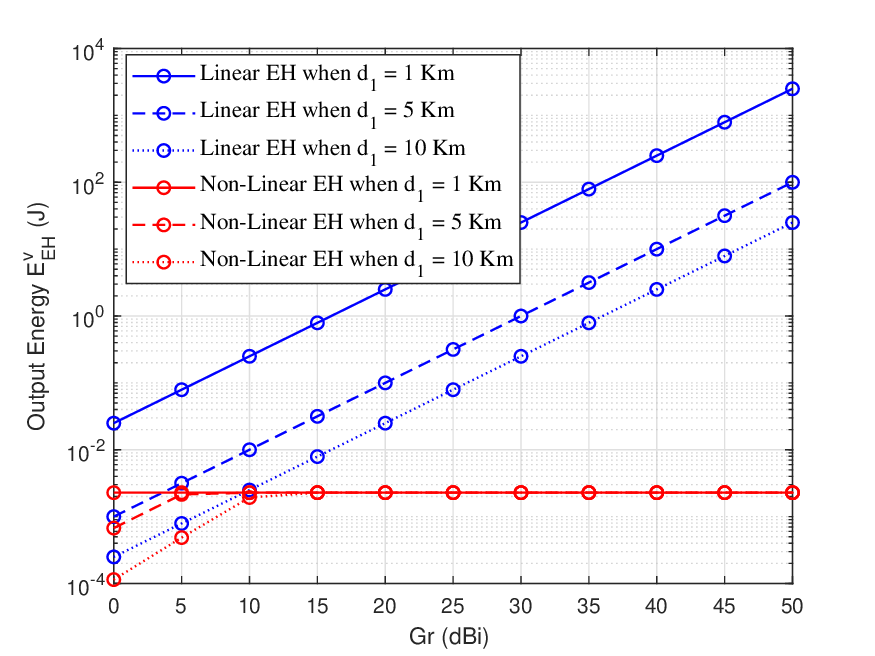}
}
\caption{Output energy $E_{\text{EH}}^{\text{v}}$ under linear and non-linear models: (a) $E_{\text{EH}}^{\text{v}}$ w.r.t. transmit power with different frequencies ($f$); (b) $E_{\text{EH}}^{\text{v}}$ w.r.t. receiver antenna gain ($G_{\text{r}}$) with different distances ($d_1$).}
\end{figure}

\hl{To ensure sufficient transmission power, harvested energy must meet a required threshold. A regular HAPS-SMBS supplements its stored energy to achieve this threshold but stops using inventory energy when harvested levels are adequate. The objective is to minimize dependence on stored energy by enhancing harvested energy, with antenna gain being one of the critical factors in this process. Therefore, for a more in-depth investigation}, Fig. 11 shows the data rate with varying receiver antenna gain when maximizing the transmit power of regular HAPS-SMBS using (35) is used when $P_{\text{Req}}=-5$ dBm. From this figure, it is evident that reducing receiver antenna gain decreases the data rate performance of the EH system. This figure grants an important observation, which is as follows: increasing receiver antenna gain indicates that using multiple antennas significantly improves system performance. 
In this figure, we also observe that the data rate of the system without EH is missing at lower transmit power levels compared to EH systems. This is because the existing transmit power at regular HAPS-SMBS is insufficient for transmission in the system without EH at those levels. 
These results provide insight into the importance of good and optimal design to implement EH in HAPS-SMBS by carefully determining antenna gains, the number of antennas, frequency, and distances.

\begin{figure}[]
    \centering
    \includegraphics[width=1\columnwidth]{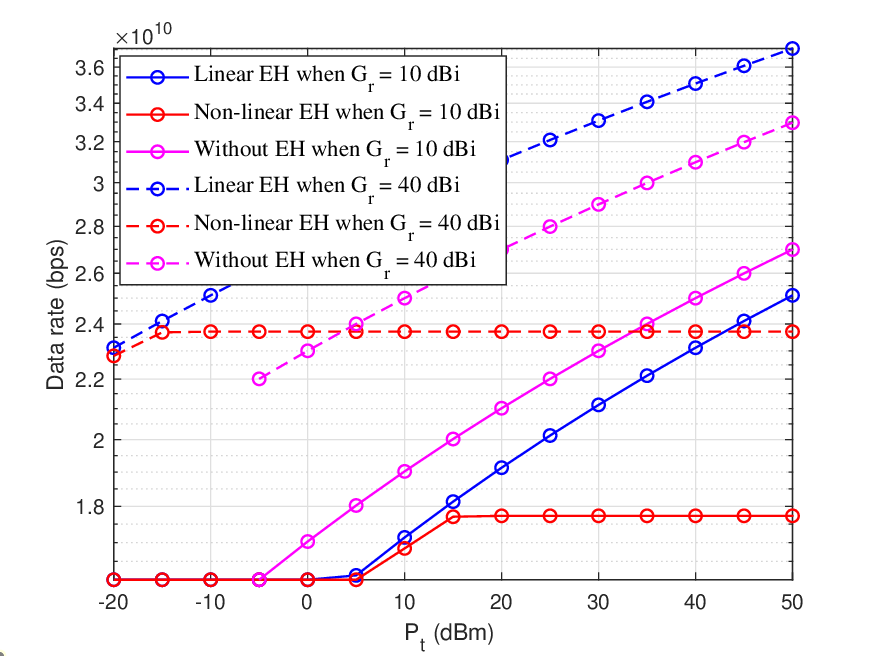}
    \caption{Data Rate comparison for linear and non-linear EH with varying receiver antenna gain ($G_{\text{r}}$).}
    \label{constellations}
\end{figure}

On the other hand, to evaluate the amount of harvested energy during the regular HAPS-SMBS flight time, in Fig. 12 (a), we present the amount of EH w.r.t. $f$ during various flight times of regular HAPS-SMBS, when $P_{\text{t}}=30$ dBm. Again, it is seen that the higher frequency reduces the amount of harvested energy. However, the increasing flight time increases the amount of the harvested energy. It is also observed that the amount of EH reduces faster in the linear EH compared to the non-linear EH as the frequency increases.
To follow the amount of harvested energy, we present in Fig. 12 (b) the amount of harvested energy w.r.t. $d_1$ (Km) during different flight times of regular HAPS-SMBS, when $P_{\text{t}}=30$ dBm. 
We observe that increasing the distance limits the amount of harvested energy.  However, the increasing flight time increases the amount of the harvested energy. It is also observed that the amount of EH reduces faster in the linear EH compared to the non-linear EH as the distance increases. 
Based on the literature, the regular HAPS-SMBS can remain in the air for a few days or months (e.g., Stratobus).  Consequently, with improved antenna designs and the integration of technologies, such as RIS and FSO, aerial platforms might be able to collect enough energy from each other during their flight time to continue flying without needing to recharge, or in some cases, extend their flight duration by a few hours.

\begin{figure}[]
\centering
\subfloat[]{%
\includegraphics[width=1\columnwidth]{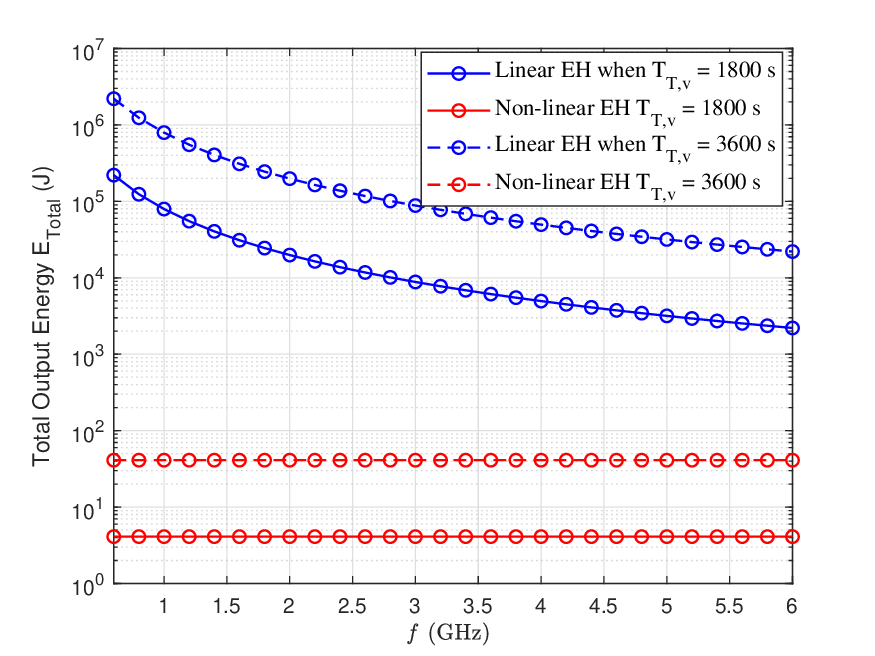}
}
\hfill
\subfloat[]{%
\includegraphics[width=1\columnwidth]{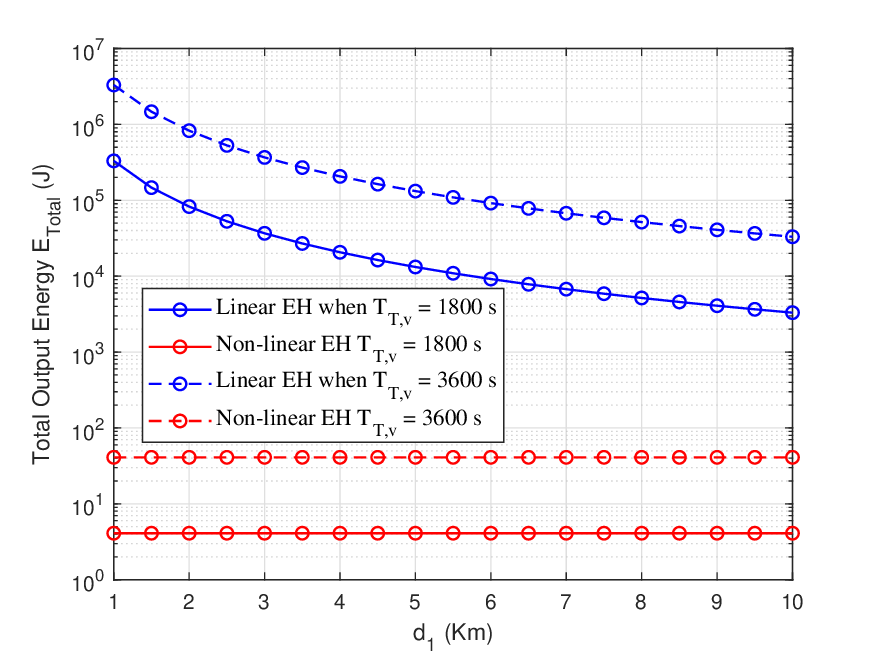}
}
\caption{Total output energy under linear and non-linear models with different flight times: (a) w.r.t. frequency ($f$); (b) w.r.t. distance ($d_{1}$ (Km)).}
\end{figure}

\section{Conclusion}
In this paper, we proposed a new paradigm for aerial platforms empowered by EH, where these platforms collect energy from signals transmitted by nearby aerial platforms.
In the proposed system, we use a two-tier HAPS-SMBS system containing regular HAPS-SMBS nodes serving as base stations, alongside a mother HAPS-SMBS node acting as a manager to coordinate and manage communications between regular HAPS-SMBS and the ground station, thus enabling wireless energy transfer.
We investigated the propagation model of HAPS-SMBS empowered by the EH for both linear and non-linear EH models. The data rate of the proposed system was obtained. Furthermore, we determined the optimal positioning for regular HAPS-SMBS to reduce signal and power loss. In addition, $Q$-learning and IDFA were employed and compared as sub-optimal solutions to achieve the joint optimization of the EH factor and regular HAPS-SMBS positioning. We proposed transmit power maximization when the minimum regular HAPS-SMBS energy is used. We discussed the HAPS-SMBS empowered by EH and compared our results to those without EH. We also assessed the amount of EH using different parameters and discussed its effectiveness. The findings revealed that the $Q$-learning and IDFA provide higher data rate performance compared to conventional EH systems. In contrast, $Q$-learning provides significant approximations in accuracy to optimal solutions in linear models with intensive training compared to IDFA.
Maximizing transmit power leads to higher efficiency gains than the system without EH. 
Furthermore, non-linear EH negatively impacts system performance, which presents real-world scenarios, necessitating further studies to mitigate it. This work will open a new challenge for researchers in the future to enhance EH in aerial platforms, such as advanced massive MIMO, RIS, FSO, and machine learning~technologies.

\section*{Acknowledgment}
This study is supported in part by the Study in Canada Scholarship (SICS) by Global Affairs Canada, in part by The Scientific and Technological Research Council of Türkiye (TUBITAK), and in part by the Discovery Grant RGPIN-2022-05231 from the Natural Sciences and Engineering Research Council of Canada (NSERC) and in part by King Abdulaziz University, Saudi Arabia.


\bibliographystyle{IEEEtran}
\bibliography{references}

\vfill

\end{document}